\documentclass{aa} %

\usepackage{graphicx}
\usepackage{txfonts}

\newcommand{\markmenew}[1]{#1}
\newcommand{\markme}[1]{#1}
\newcommand{\srem}[1]{}
\usepackage{psfrag,pstricks,epsfig,tabularx}
\usepackage{epstopdf}
\usepackage{ulem}

\begin{document} 

\title{Evolution of an equatorial coronal hole structure and the released coronal hole wind stream: \markme{Carrington rotations 2039 to 2050}}
\titlerunning{Evolution of an equatorial coronal hole structure and wind}
\author{Verena Heidrich-Meisner \inst{1}, Thies Peleikis \inst{1}, Martin Kruse \inst{1}, Lars Berger \inst{1}, and Robert F. Wimmer-Schweingruber \inst{1}}

\authorrunning{Heidrich-Meisner et al.} 

\institute{Christian Albrechts University at Kiel, Germany,
  \email{heidrich@physik.uni-kiel.de}
}

\date{}

\abstract{
The Sun is a highly dynamic environment that exhibits dynamic behavior on many different timescales. \markme{Variability is observed both in closed and in open field line regions in the solar corona. In particular, coronal holes exhibit temporal and spatial variability. Signatures of these coronal dynamics are inherited by the coronal hole wind streams that originate in these regions and can effect the Earth's magnetosphere. Both the cause of the observed variabilities and how these translate to fluctuations in the in situ observed solar wind is not yet fully understood. }
}{
  During solar activity minimum the structure of the magnetic field typically remains stable over several Carrington rotations (CRs). But how stable is the solar magnetic field? Here, we address this question by analyzing the evolution of a coronal hole structure and the corresponding coronal hole wind stream emitted from this source region over 12 consecutive CRs in 2006. 
}{
  To this end, we link in situ observations of Solar Wind Ion Composition Spectrometer (SWICS) onboard the Advanced Composition Explorer (ACE) with synoptic maps of Michelson Doppler imager (MDI) on the Solar and Heliospheric Observatory (SOHO) at the photospheric level through a combination of ballistic back-mapping and a potential field source surface (PFSS) approach. Together, these track the evolution of the open field line region that is identified as the source region of a recurring coronal hole wind stream.
  Under the assumptions of the freeze-in scenario for charge states in the solar wind, we derive freeze-in temperatures and determine the order in which the different charge state ratios of ion pairs appear to freeze-in. We call the combination of freeze-in temperatures derived from in situ observed ion density ratios and freeze-in order a minimal electron temperature profile and investigate its variability. 
}{
  The in situ properties and the PFSS model together probe the lateral magnetic field configuration, the minimal temperature profiles allow to constrain the radial structure. 
    We find that the shape of the open field line region and to some extent also the solar wind properties are influenced by surrounding more dynamic closed loop regions.  We show that the freeze-in order can change within a coronal hole wind stream on small timescales and illustrate a mechanism that can cause changes in the freeze-in order. The inferred minimal temperature profile is variable even within coronal hole wind and is in particular most variable in the outer corona.
  
}


   \maketitle
%

\section{Introduction}
In situ observations of the solar wind often show reoccurring structures that are interpreted as the footprint of stable coronal structures at the Sun. This is supported by remote sensing observations that show coronal structures that remain stable over several Carrington rotations (CRs) \citep[]{benevolenskaya2001active,russell2010unprecedented}. These stable structures are typically observed  during solar minimum conditions. Under these conditions, when the Sun is less active, the solar magnetic field changes less rapidly than during the solar activity maximum; fewer active regions, fewer sunspots, and from a  heliospheric point of view most importantly fewer coronal mass ejections (CMEs) occur during solar minimum than during solar maximum \citep[]{gopalswamy2006coronal,robbrecht2009automated}. The quiet Sun leads to a less variable heliospheric structure as well. But the Sun is never completely quiet even during deep solar minimum. Although the large scale changes occur more slowly, the Sun is nevertheless still a highly dynamic environment and the magnetic configuration continues to evolve during solar minimum. A recurring stable structure on the Sun can be considered as a test bed to investigate the comparatively slow evolution of this region and the solar wind streaming into the heliosphere from this source region.
Except for the influence of wave activity and  stream interaction regions, the fine structure of in situ observed solar wind properties can be linked back to the evolving conditions at the solar source regions. The solar wind speed, temperature, ion, and charge state composition, as well as the magnetic field are all influenced by the conditions in the respective source region of a solar wind \markme{stream.}

\markme{The} average magnetic field configuration at the photospheric level is captured by synoptic maps based on magnetograms from, for example, the Michelson Doppler imager \markme{\citep[MDI, ][]{scherrer1991solar}} on the Solar and Heliospheric Observatory (SOHO). The corresponding coronal structure of the magnetic field can be derived with a potential field source surface approach \markme{\citep[PFSS, ][]{schatten-etal-1969,altschuler-and-newkirk-1969}}. This procedure requires the magnetic configuration on the complete solar surface as an input, such as MDI synoptic maps. These combine observations from a complete CR  into an average map. As a result the derived magnetic field configuration cannot reflect dynamic changes on timescales faster  than a CR and is best applied during solar minimum conditions. This limits the eligibility of PFSS models. However, in this case study, we are especially interested in the scenario PFSS models are best suited for: We are looking for changing fine structure features in the photosphere, solar corona, and in situ observations that vary on the timescale of \markme{CRs.} 

\markme{The} PFSS model enables us to track magnetic field lines from the photosphere to the source surface which is conventionally but arbitrarily assumed to be at $2.5$ solar radii ($R_\odot$). Under the assumption that the solar wind speed is constant from the source surface to the location of the spacecraft, in our case the Advanced Composition Explorer (ACE) at L1, a solar wind package can be tracked ballistically from the spacecraft to the source surface. However, stream interaction regions obviously violate the assumption of constant solar wind speed. Moreover, especially the fast solar wind is likely to experience ongoing acceleration much further out in the corona \citep[]{cranmer2009coronal,fisk1999acceleration}. This type of combination of ballistic back-mapping and the PFSS model has frequently been  used to investigate the source regions of solar wind \citep[]{fazakerley2016investigation,gomez2011august,thompson2011snapshot}.

The combination of in situ observations and the back-mapping method enables us to trace the fine structure at the photospheric level for each CR. The comparison between CRs enables us to track the temporal evolution of these structures. This can be considered as a lateral (or horizontal) probing of the evolution of magnetic structures in the photosphere.  The aim of this study is to investigate variability inside coronal hole wind flows.  Thus, we focus mainly on one recurring coronal hole wind stream per CR. This stream has another interesting property with respect to its Fe charge state composition. \citet[]{heidrich2016observations} has shown that coronal hole wind streams with either low Fe charge states (Fe-cool) or Fe charge states comparable to those observed in slow solar wind (Fe-hot) are observed in equatorial coronal hole wind. In particular, transitions from Fe-cool to Fe-hot coronal hole wind or vice versa were observed. Our coronal hole wind stream of interest shows  transitions between Fe-hot and Fe-cool coronal hole wind. In the context of this investigation, the average Fe charge state thus provides an additional tool to investigate the fine structure of a coronal hole wind stream.

Tracking the radial evolution of the magnetic field line configuration in the solar atmosphere is more difficult. Without invoking a more or less complete model of the solar atmosphere it is, for example, not possible to derive a radial temperature profile of the solar atmosphere solely from in situ observations. However, we argue that even without an additional model, in situ derived freeze-in temperatures  \markme{\citep[]{hundhausen1968state,ogilvie1980variation,owocki1983solar,geiss1995}} can give some insights on radial structures and their variability.
In the freeze-in, image charge states can change freely as long as the charge modification timescale is smaller than the expansion timescale. For different ion species this condition is met at different heights in the solar atmosphere. In particular, the order in which the charge states of each ion pair freezes is constrained by the local electron density and - through the recombination and ionization rates - by the electron temperature. The charge modification timescale is sensitive to the relative influence of ionization compared to recombination. This can also change the order in which ion pairs freeze-in, as  illustrated in Sect.~\ref{sec:vertical}.    
We combine the information on the order in which the charge states freeze-in with the freeze-in temperatures into what we call a minimal temperature profile. These minimal temperature profiles can be considered as a tracer of radial (or vertical) structures in the solar atmosphere. Their temporal evolution can provide insights on the variability and radial evolution of coronal structures without requiring a complete model of the solar atmosphere.

This article is structured as follows: Section~\ref{sec:data} describes our data selection and the back-mapping procedure. In Sect.~\ref{sec:horizontal}, we investigate the variability of our coronal structure of interest from a photospheric and in situ perspective. We then discuss minimal temperature profiles and their variability in Sect.~\ref{sec:vertical}. In Sect.~\ref{sec:conclusions}, we present our conclusions.

\section{Data selection, solar wind characterisation, and methods}\label{sec:data}
We combine in situ observations from ACE/SWICS \citep[]{gloeckler-etal-1998}, ACE/MAG \citep[]{smith1998ace} and the Solar Wind Electron, Proton and Alpha Monitor (ACE/SWEPAM) \citep[]{mccomas1998solar} with synoptic maps based on magnetograms from SOHO/MDI \citep[]{scherrer1991solar}. The analysis procedure applied to the raw ACE/SWICS data is described in detail in \citet[]{berger2008velocity} and has been applied in, for example, \citet[]{berger2011systematic} and \citet[]{heidrich2016observations}.

\begin{figure*}
\includegraphics[width=\textwidth]{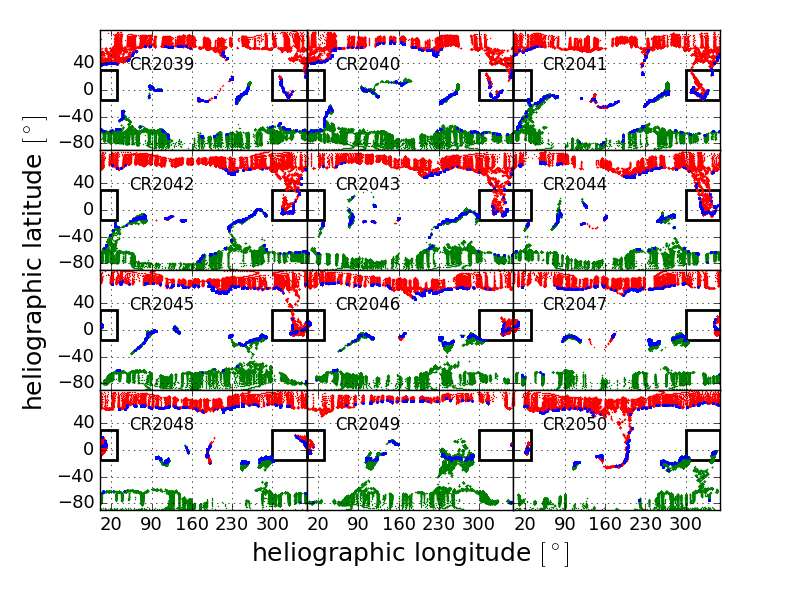}
\caption{\label{fig:CRmap} Heliographic map at the photospheric level for CRs \markme{2039-2050} derived from a numerical PFSS approach. The maps are sorted row-wise by CR with CR 2039 at the top left and CR 2050 at the bottom right. Only foot points of open field lines are shown. Inwards directed magnetic polarity is indicated with red dots and outwards directed polarity with green dots. The blue dots refer to the foot points of ACE mapped down to the photosphere. The black boxes indicate the region of interest for which a zoom-in is shown in the following Fig.~\ref{fig:CRmapholes}.}
\end{figure*}

\begin{figure*}
\includegraphics[width=\textwidth]{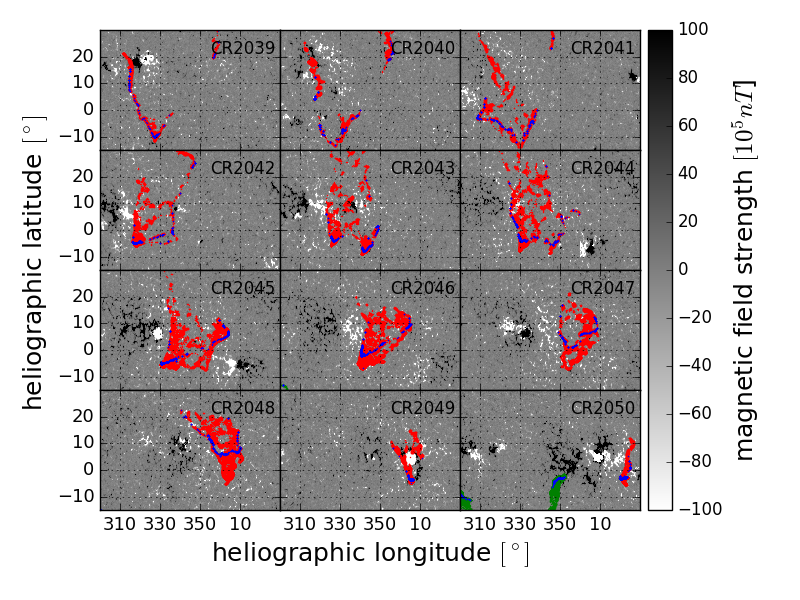}
\caption{\label{fig:CRmapholes} Zoom of the heliographic map at the photospheric level for CRs \markme{2039-2050}. In addition to the foot points of open field lines as in Fig.~\ref{fig:CRmap}, the respective MDI synoptic maps that were used as input to the PFSS model are shown in the background.}
\end{figure*}


To characterize observed solar-wind packages by type, we use the categorisation scheme from \citet[]{xu2014new}. This scheme defines four categories of solar wind: coronal hole wind plasma, ejecta plasma, and two slow solar wind categories, namely sector-reversal region plasma and streamer-belt plasma. Here, we combine the latter two into a single category for slow solar wind. The ejecta category is intended to cover plasma from interplanetary coronal mass ejections (ICMEs). However, this category can be problematic because it frequently and possibly incorrectly interprets parts of cool, very slow solar wind streams as ejecta plasma. Therefore, we instead rely on the \citet[]{jian2006properties,jian2011comparing}  ICME list for identifying ICMEs in the in situ observations. The Xu \& Borovsky scheme is applied to data with the native 12-min resolution of ACE/SWICS.

We chose ten consecutive CRs in 2006 as a case study to investigate the evolution of the source region of a recurring  coronal hole wind stream. This stream and these CRs were selected not only because the coronal hole wind stream is observed repeatedly but also because it exhibits an interesting feature with respect to its average Fe charge state \citep[]{heidrich2016observations}.  These streams show transition between coronal hole wind with low Fe charge states (called Fe-cool coronal hole wind) and coronal hole wind with high Fe charge states (Fe-hot coronal hole wind). 
The distinction between Fe-hot and Fe-cool coronal hole wind streams is based on a comparison of the average Fe-charge state $\tilde{q}_{Fe} = \sum_{c=7}^{13} c n_{Fe^{c+}}/ \sum_{c=7}^{13} n_{Fe^{c+}}$ in coronal hole wind streams to the average Fe charge state of all (pure) slow solar wind in the same year. As in \citet[]{heidrich2016observations}, to avoid both compression and rarefaction regions between slow and coronal  wind streams the proton-proton collisional age $a_c$ \markme{\citep[]{kasper2008hot,heidrich2016observations}} is used as an additional criterion for solar wind classification. In particular, an upper bound $a_c<0.1$ is applied for choosing pure coronal hole wind for the coronal hole wind streams of interest and a lower bound $a_c>0.4$ for pure slow solar wind as reference for defining the notion of high (Fe-hot) and low (Fe-cool) Fe charge states. Thus, solar wind plasma that has been categorized as coronal hole wind by the Xu \& Borovsky scheme is considered to be so-called pure coronal hole wind if, additionally, the collisional age is low, $a_c<0.1$, and solar wind plasma that has been categorized as slow solar wind in the Xu \&Borovsky scheme if considered pure slow solar wind if the collisional age is high, $a_c>0.4$. These two categories (pure coronal hole wind and pure slow solar wind, respectively) are only used for the definition of Fe-hot and Fe-cool coronal hole wind and are applied on observations with 4h time resolution. We use the modified Xu \& Borovsky scheme here instead of the categorization of slow and fast solar wind used in \citet[]{heidrich2016observations} to test how sensitive the characterization of Fe-hot or Fe-cool coronal hole wind is to the particular solar wind categorization scheme. \markme{As a result of the different underlying solar wind categorization scheme, t}he threshold value for the average Fe charge state that differentiates between Fe-cool and Fe-hot coronal hole wind changes for the year 2006 from $\tilde{q}_{Fe, slow, 2006}=9.71$ \markme{\citep[which was used in][]{heidrich2016observations}} to $\tilde{q}_{Fe, slow, 2006}=9.66$ \markme{(which is used here). Consequently,} the start and end times of each stream change slightly and, in some cases, more transition occurs. Nevertheless,  the presence of Fe-cool and Fe-hot coronal hole wind streams and the existence of transitions between them is unaffected by the particular solar wind categorization scheme. \markme{The mean of all average Fe charge states in Fe-hot coronal hole wind is during CR 2039-2050 $\tilde{q}_{Fe-hot}=9.75$ and for the same CRs in Fe-cool coronal hole wind the respective mean value is $\tilde{q}_{Fe-cool}=9.65$.}

To link the in situ observations with the photosphere, we apply a combination of ballistic back-mapping and the PFSS model \citep[]{schatten-etal-1969,altschuler-and-newkirk-1969}. The same approach has been applied in \citet[]{heidrich2016observations} and \citet[]{peleikis2015sw14,peleikis2017}. Firstly, a solar wind package is mapped back ballistically to the source surface at $2.5 R_\odot$ under the assumption that the solar wind speed is constant between the source surface and the observer. Secondly, a grid of open field lines is distributed over the source surface. The resolution in heliographic latitude is latitude-dependent with a finer grid at the equator ($0.63 ^\circ$) than in polar regions ($2.2^\circ$). The longitudinal grid is uniform with $1^\circ$ resolution. Thirdly, based on a numerical PFSS approach on this grid with a 50 radial grid points, each open field line is traced down to the photosphere. The PFSS model requires a synoptic map of the complete photospheric surface as a boundary condition. This limits the applicability of this approach. The PFSS model is most reliable during solar minimum when the magnetic field configuration changes only slowly and few or no CMEs are observed. However, this is exactly the situation we are interested in in this case. Thus, the influence of averaging the solar magnetic field over a complete CR still limits the accuracy of the back-mapping but less so than under different conditions. \markme{The assumption of a constant solar wind speed that is required by the ballistic back-mapping is invalid for stream interaction regions. As a result, the back-mapping is distorted for stream-interaction regions. This effect can be reduced by the upwind back-mapping algorithm in \citet[]{riley2011mapping}. However, since we are mainly interested in coronal hole wind streams, the upwind back-mapping algorithm was not applied here.}

\begin{figure*}  \begin{center}
\includegraphics[width=.8\textwidth]{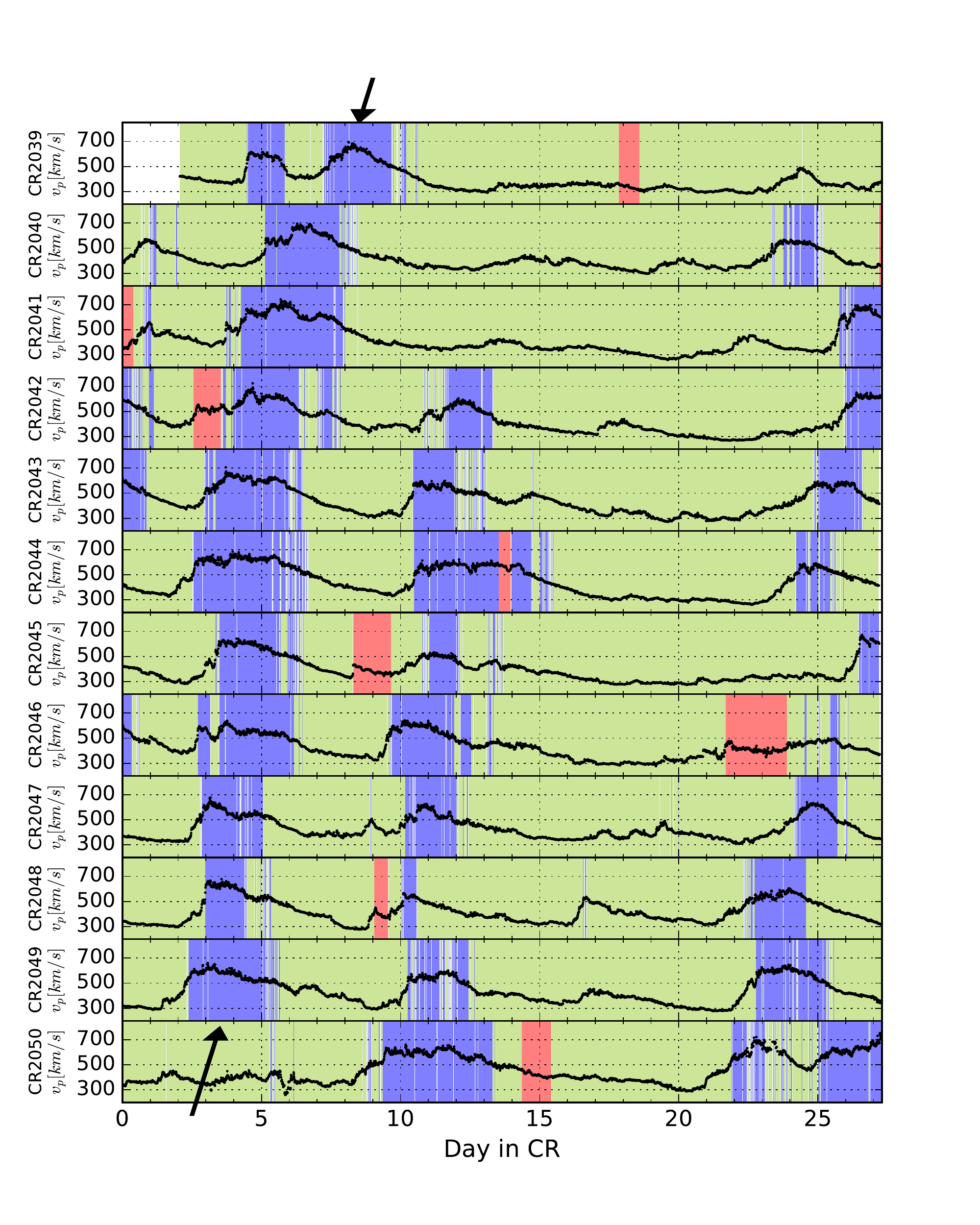}
  \end{center}
  
\caption{\label{fig:CRvsw} Solar wind speed over time for CRs \markme{2039-2050}. Coronal hole wind is highlighted in blue and slow solar wind in green based on the Xu \& Burovsky scheme. ICMEs from the Jian ICME catalog are indicated in red. The arrows indicate our coronal hole wind stream of interest.}
\end{figure*}

\begin{figure*}
  \begin{center}
    \includegraphics[width=.9\textwidth]{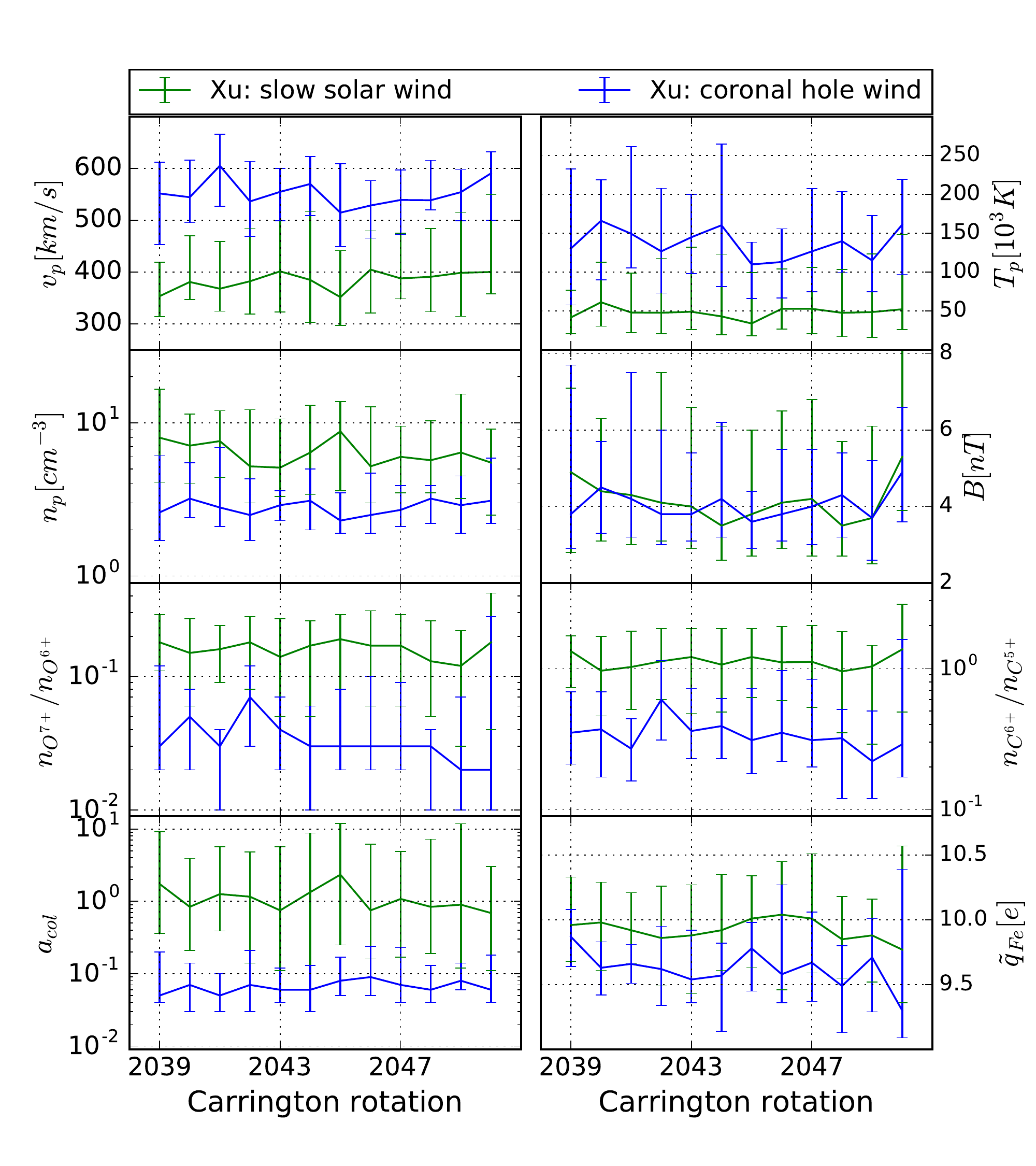}

  \end{center}
    \caption{\label{fig:properties} Solar wind properties per solar wind type based on the Xu \& Burovsky scheme (coronal hole wind (blue) and slow solar wind (green)), averaged over the respective CRs. In each panel, the median of all 4h bins of the respective solar wind type is shown, as well as the true variability of the 4h-resolution observations as indicated by the \markme{15.9th and 84.1th }percentiles. Row-wise from top left to bottom right: proton solar wind speed $v_p$, proton temperature $T_p$, proton density $n_p$,  magnetic field strength $B$, O charge state ratio  $n_{O^{7+}}/n_{O^{6+}}$, C charge state ratio $n_{C^{6+}}/n_{C^{5+}}$, collisional age $a_c$, and average Fe charge state $\tilde{q}_{Fe}$.}
\end{figure*}

\begin{figure*}  \begin{center}
\includegraphics[width=.8\textwidth]{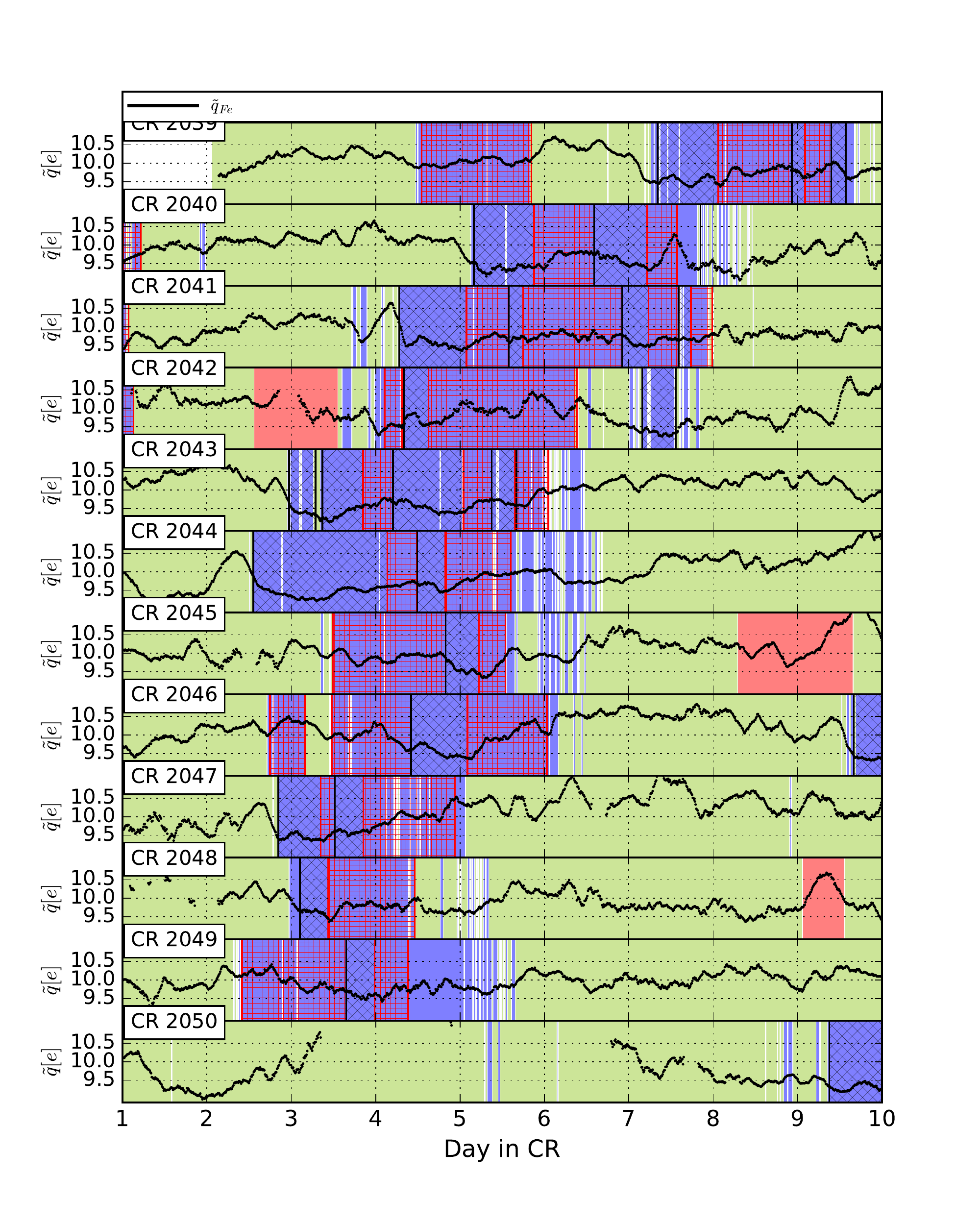}
  \end{center}
  
  \caption{\label{fig:CRFe} Average Fe charge state over time for CRs \markme{2039-2050}. The solar wind type is indicated in the same way as in Fig.~\ref{fig:CRvsw}. Periods with high and low Fe charge states in coronal hole wind are highlighted:  Fe-hot (red, $+$-hatched) and Fe-cool (black, $x$-hatched).}
\end{figure*}

\begin{figure*}  \begin{center}
\includegraphics[width=0.8\textwidth]{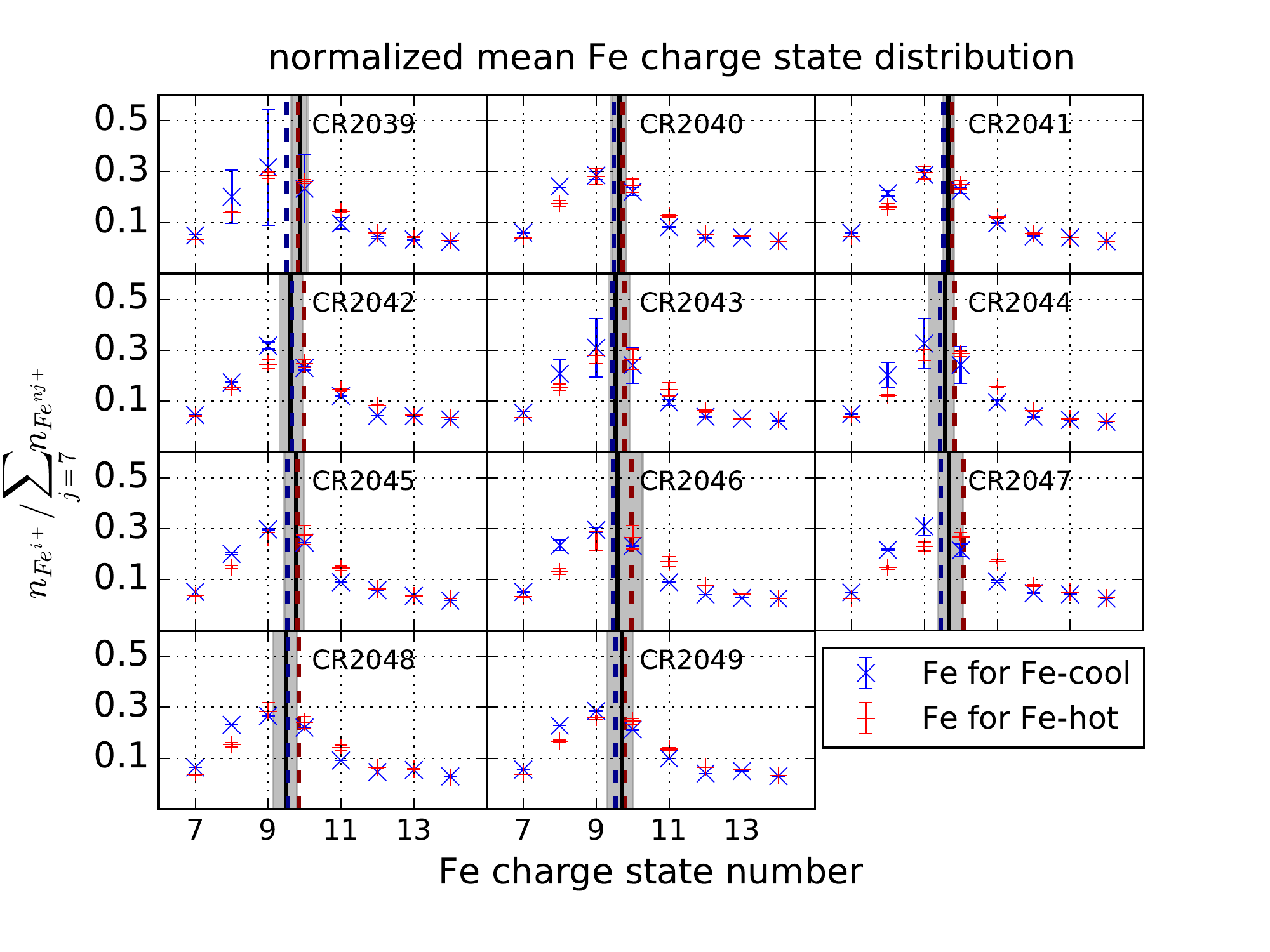}
  \end{center}
  
\caption{\label{fig:CSCompFe} Normalized mean charge distributions for Fe sorted by CR (in the same order as in Fig.~\ref{fig:CRmap}) averaged separately over the Fe-hot and Fe-cool parts of the first recurring coronal hole stream during CR 2030-2049. CR 2050 is omitted because the coronal hole wind stream is not observed any more in that CR. The black vertical line marks the median of the  average Fe charge state for each CR as given in Fig.~\ref{fig:properties} and the gray shaded areas indicate the corresponding 1$\sigma$ interval. The median of the average Fe charge state in Fe-hot is indicated by a red \markme{dashed} vertical line and the median of the average Fe charge state in Fe-cool coronal hole wind is given in blue. The error bars indicate the error of the mean. In the cases where the error bars are not visible, they are hidden behind the respective symbol for the mean. Each distribution is normalized to its sum.}
\end{figure*}

\begin{figure*}  \begin{center}
    \includegraphics[width=0.8\textwidth]{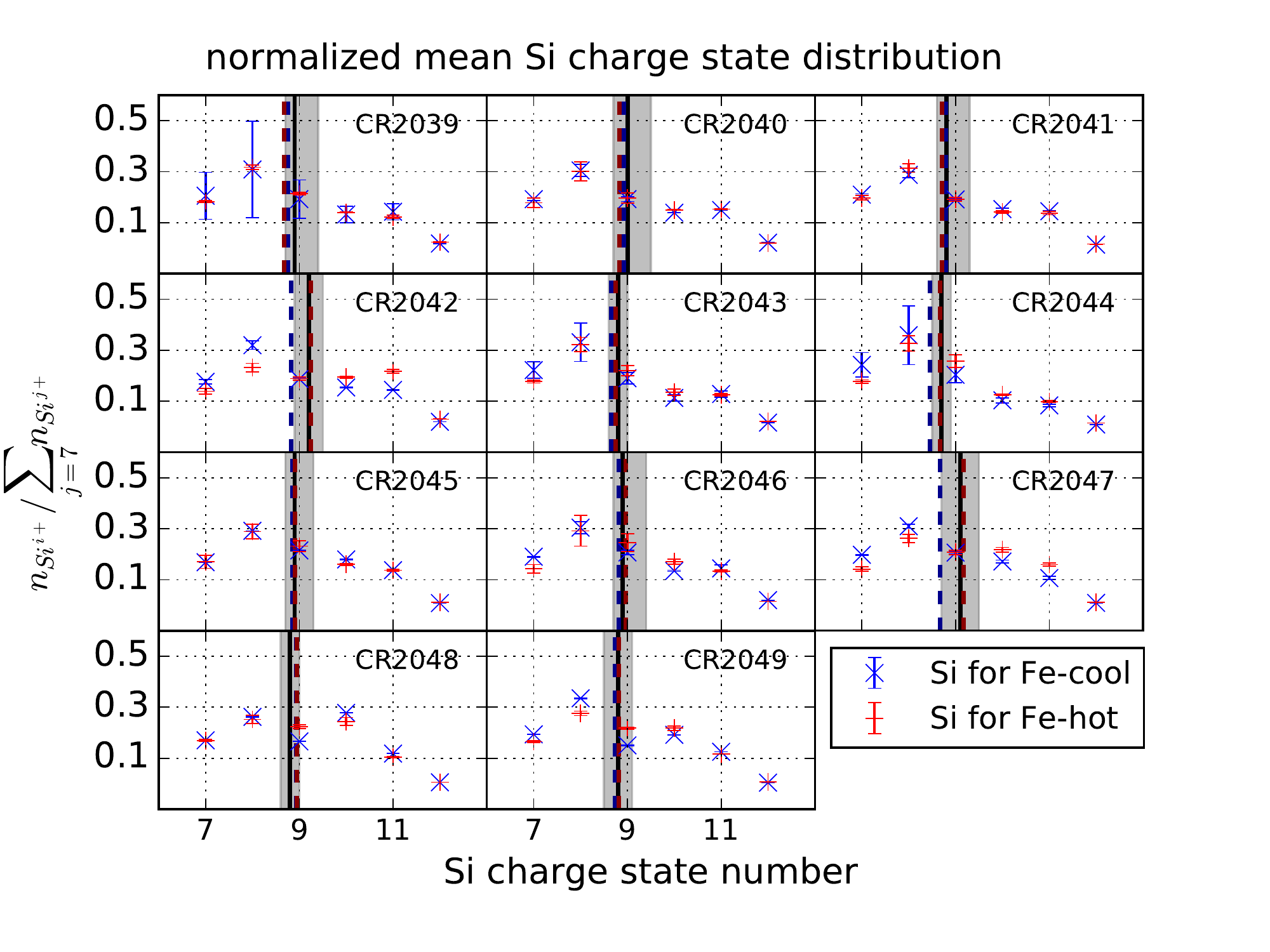}
  \end{center}
  
\caption{\label{fig:CSCompSi} Normalized mean charge distributions for Si sorted by CR (in the same order and format as in Fig.~\ref{fig:CRmap}) for the Fe-hot and Fe-cool parts of the first recurring coronal hole stream during CR 2030-2049.}
\end{figure*}

\begin{figure}
  \begin{center}

\includegraphics[width=\columnwidth]{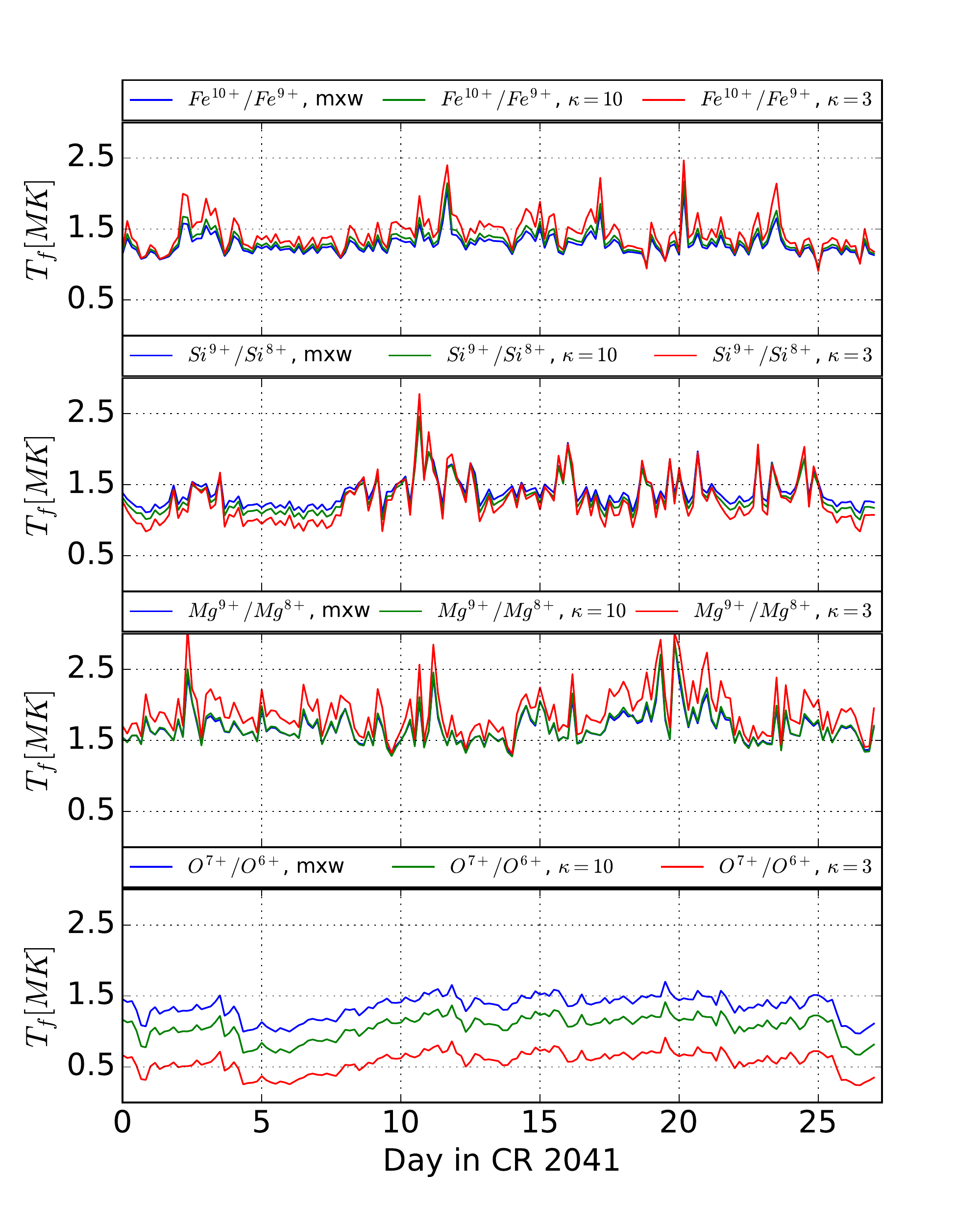}
  \end{center}

\caption{\label{fig:Tfmxwkappa} Time series of freeze-in temperatures for selected ion pairs ($ T_{f, Fe^{10+}/Fe^{9+}},  T_{f, Si^{9+}/Si^{8+}}, T_{f, Mg^{9+}/Mg^{8+}},$ and $T_{f, O^{7+}/O^{6+}}$) based on different electron velocity distribution functions in 4-hour time resolution. In each panel the freeze-in temperatures based on a Maxwellian (mxw, blue), $\kappa$-distribution with $\kappa=10$ (green), and $\kappa$-distribution with $\kappa=3$ (red) as electron velocity distribution function are shown. }
\end{figure}

\begin{figure*}
  \includegraphics[width=\textwidth]{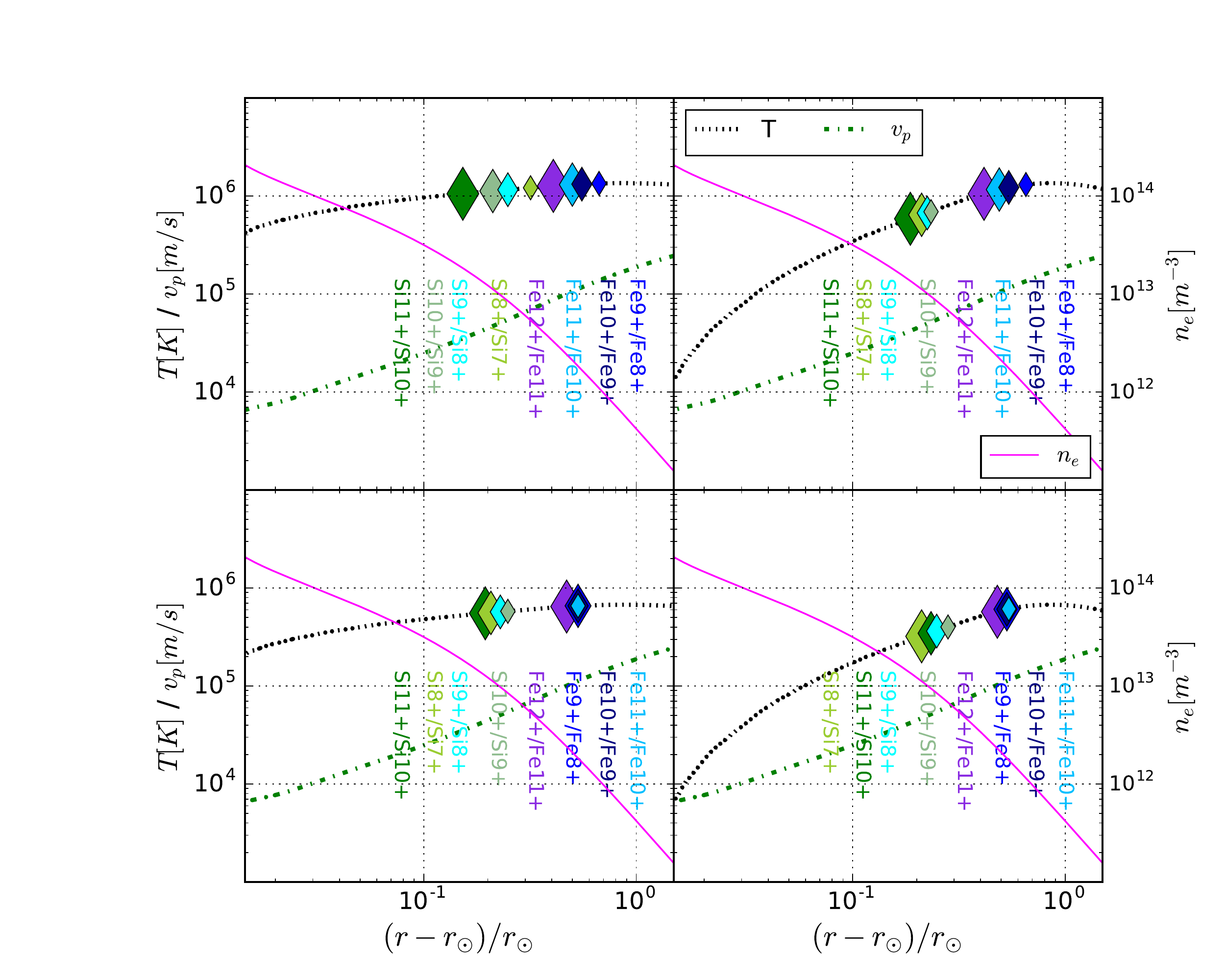}
  \caption{\label{fig:profiles} Freeze-in temperatures based on the \citet[]{cranmer2007self} model. Each panel shows  the top the temperature, electron density and bulk solar wind flow speed over the radius. \markmenew{Four} different temperature profiles are considered: unmodified (top right),  steeper increasing and declining slopes (top right), lower maximum temperature (bottom \markmenew{left}), lower maximum temperature and steeper slopes (bottom right). The freeze-in points were derived based on ionization and recombination rates for $\kappa$-distributed electron velocity functions with $\kappa=10$.}
\end{figure*}

\begin{figure*}
  \includegraphics[width=\textwidth]{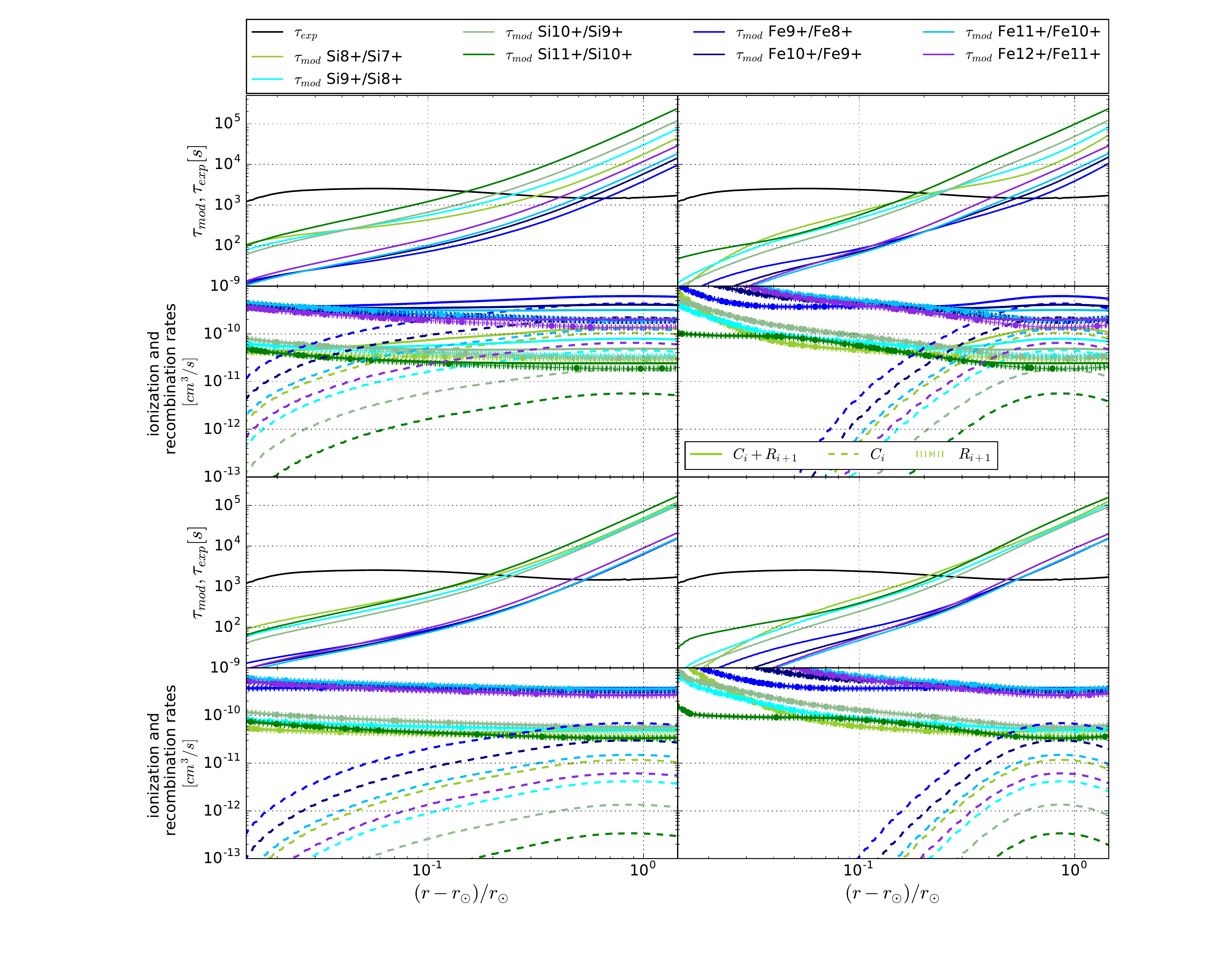}

  \caption{\label{fig:timescales} Charge modification and expansion timescales for Si and Fe ions based on the \citet[]{cranmer2007self} model (top right) and for modified temperature profiles:  steeper increasing and declining slopes (top right), lower maximum temperature (bottom \markmenew{left}), lower maximum temperature and steeper slopes (bottom right). The first and third panel row shows the expansion and charge modification timescales. The other panels show for each ion pair the ionization and recombination rates, as well as their sum. The ionization and recombination rates are based on $\kappa$-distributed electron velocity functions with $\kappa=10$.}
\end{figure*}

\begin{figure}
  \includegraphics[width=\columnwidth]{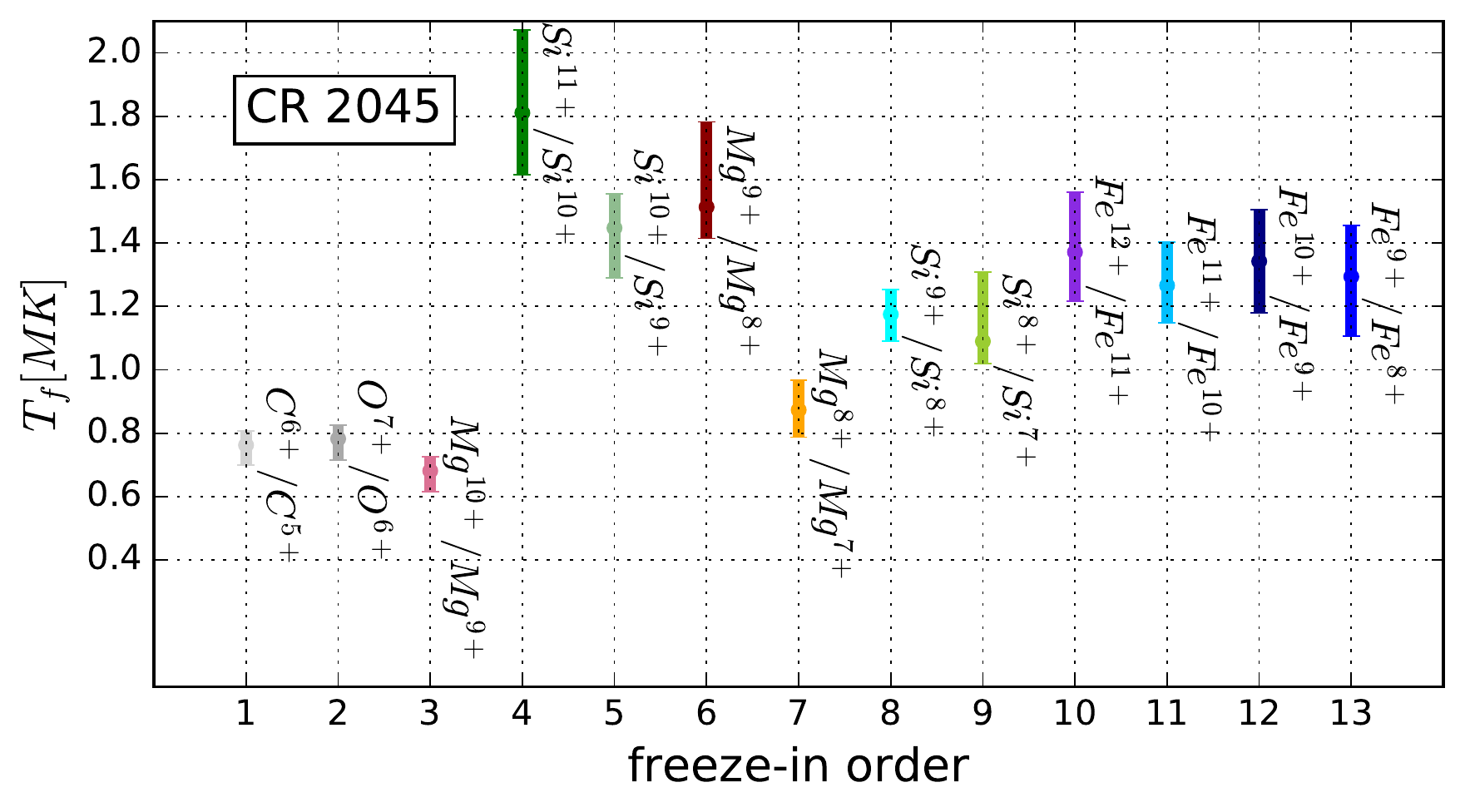}
  \caption{\label{fig:tfmedian} \markme{Minimal temperature profile averaged over the first coronal hole stream in CR 2045. The freeze-in temperatures are derived from in situ observed ion ratios and the median over the coronal hole stream is shown together with the 15.9th and 84.1th percentile as error bars.  The freeze-in order is determined by the median of the recombination and ionization rates $C_{i}(T_f) + R_{i+1}(T_f)$ at the freeze-in points. The recombination and ionization rates are based on a $\kappa$-function as electron velocity distribution with $\kappa=10$.}}
\end{figure}

\begin{figure}  \begin{center}

  \includegraphics[width=0.9\columnwidth]{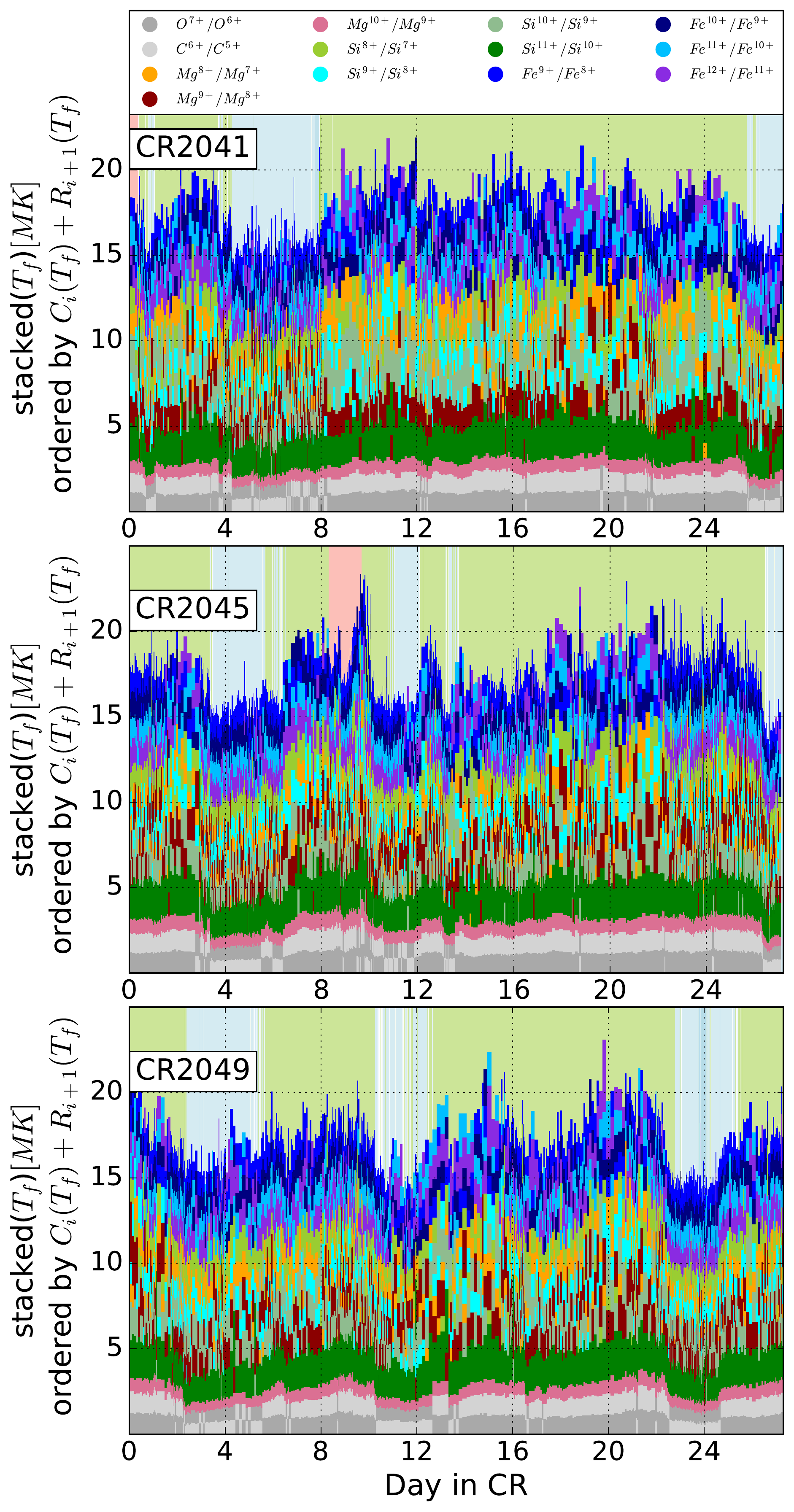}
  \end{center}

\caption{\label{fig:TfTCR} Minimal temperature profiles: Freeze-in temperatures for selected ion pairs ordered by the sum of recombination and ionization rates and over time for CR 2041 (top), CR 2045 (middle) and CR 2049 (bottom). The bin size is adaptive to ensure comparable statistics in each bin. For each element, at least 100 total counts are contained per bin. The maximum bin size is 4 hours. All panels refer to a $\kappa$-function with $\kappa=10$ as the electron velocity distribution function. Each ion pair is identified by color. Their order (from bottom to top) is determined by their recombination and ionization rates $C_{i}(T_f) + R_{i+1}(T_f)$.}
\end{figure}

\begin{figure*}
  \begin{center}
  \includegraphics[width=\textwidth]{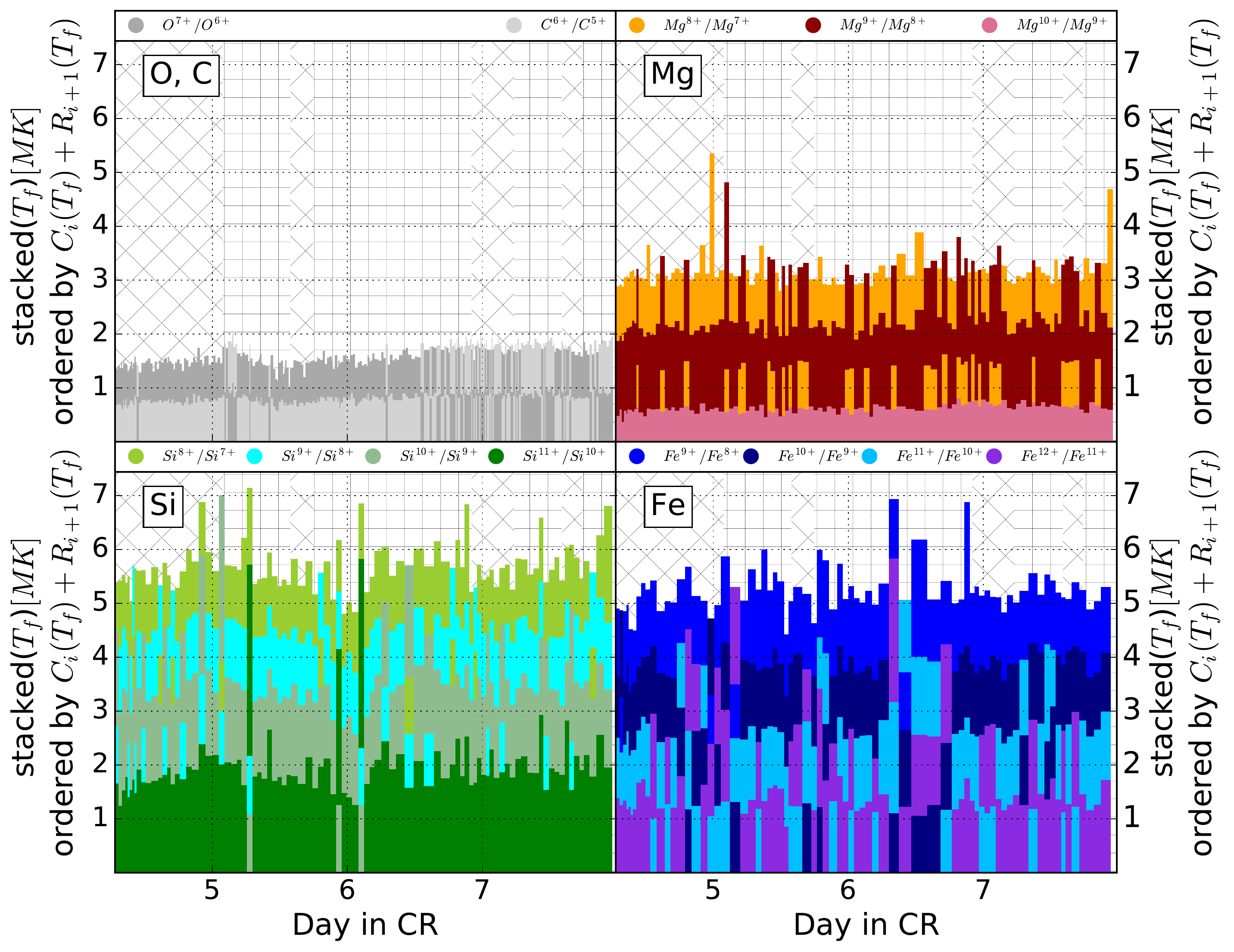}

  \end{center}

  \caption{\label{fig:TfTCR041} Time series of minimal temperature profiles with adaptive bin size for the first recurring coronal hole wind stream in CR 2041 and in the same format as in Fig.~\ref{fig:TfTCR}. Top left: O and C ion pairs, top right: Mg ion pairs, bottom left: Si ion pairs and, bottom right: Fe ion pairs. All panels refer to a $\kappa$-function with $\kappa=10$ as the electron velocity distribution function. Fe-hot and Fe-cool coronal hole wind streams are marked with $+$-shaped hatching (Fe-hot) and $x$-shaped hatching (Fe-cool).}
\end{figure*}

\begin{figure*}
    \begin{center}
  \includegraphics[width=\textwidth]{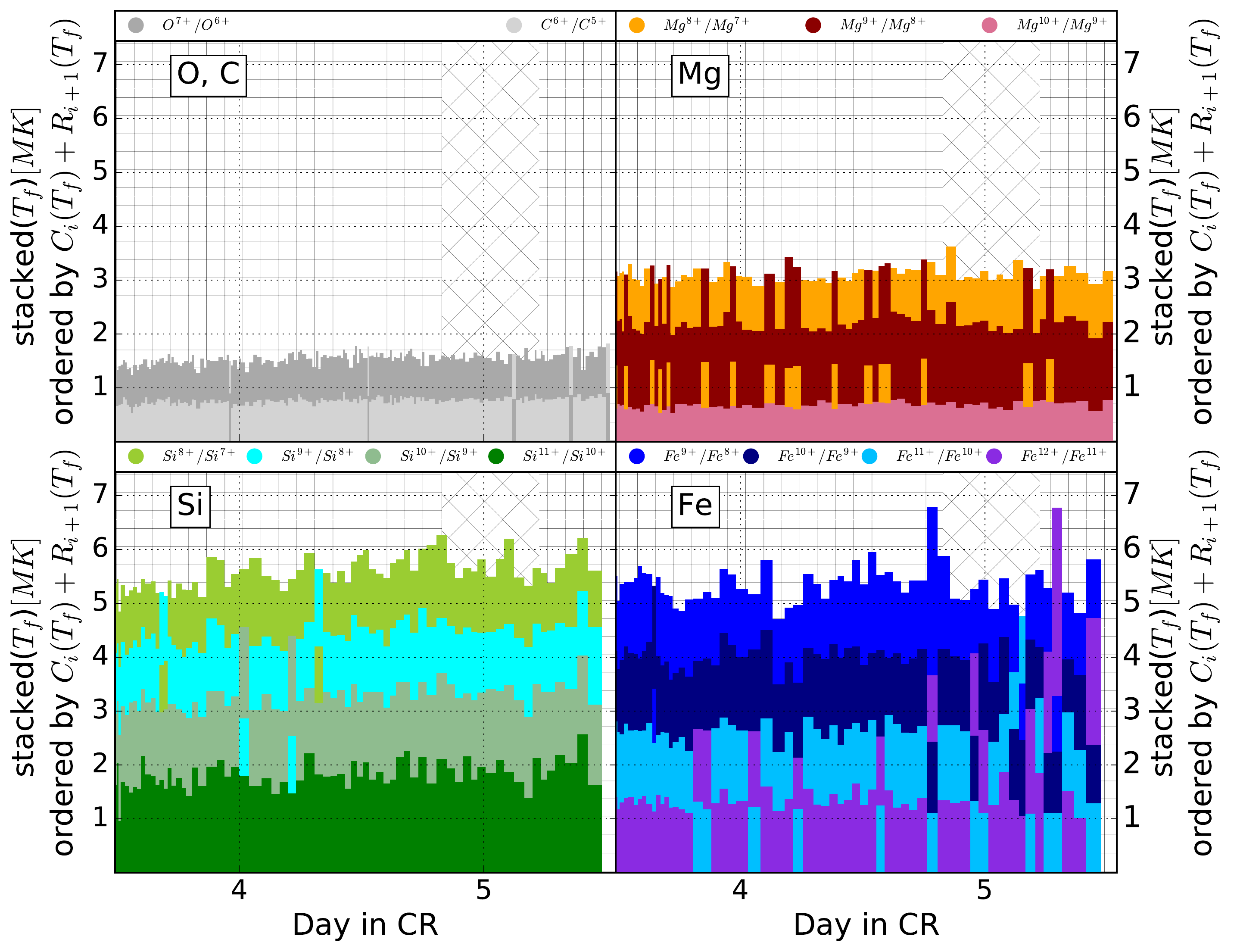}

    \end{center}

\caption{\label{fig:TfTCR045}  Time series of minimal temperature profiles with adaptive bin size for the first recurring coronal hole wind stream in CR 2045 and in the same format as in Fig.~\ref{fig:TfTCR041}. Top left: O and C ion pairs, top right: Mg ion pairs, bottom left: Si ion pairs and, bottom right: Fe ion pairs. All panels refer to a $\kappa$-function with $\kappa=10$ as the electron velocity distribution function.}
\end{figure*}

\begin{figure*}
  \includegraphics[width=\textwidth]{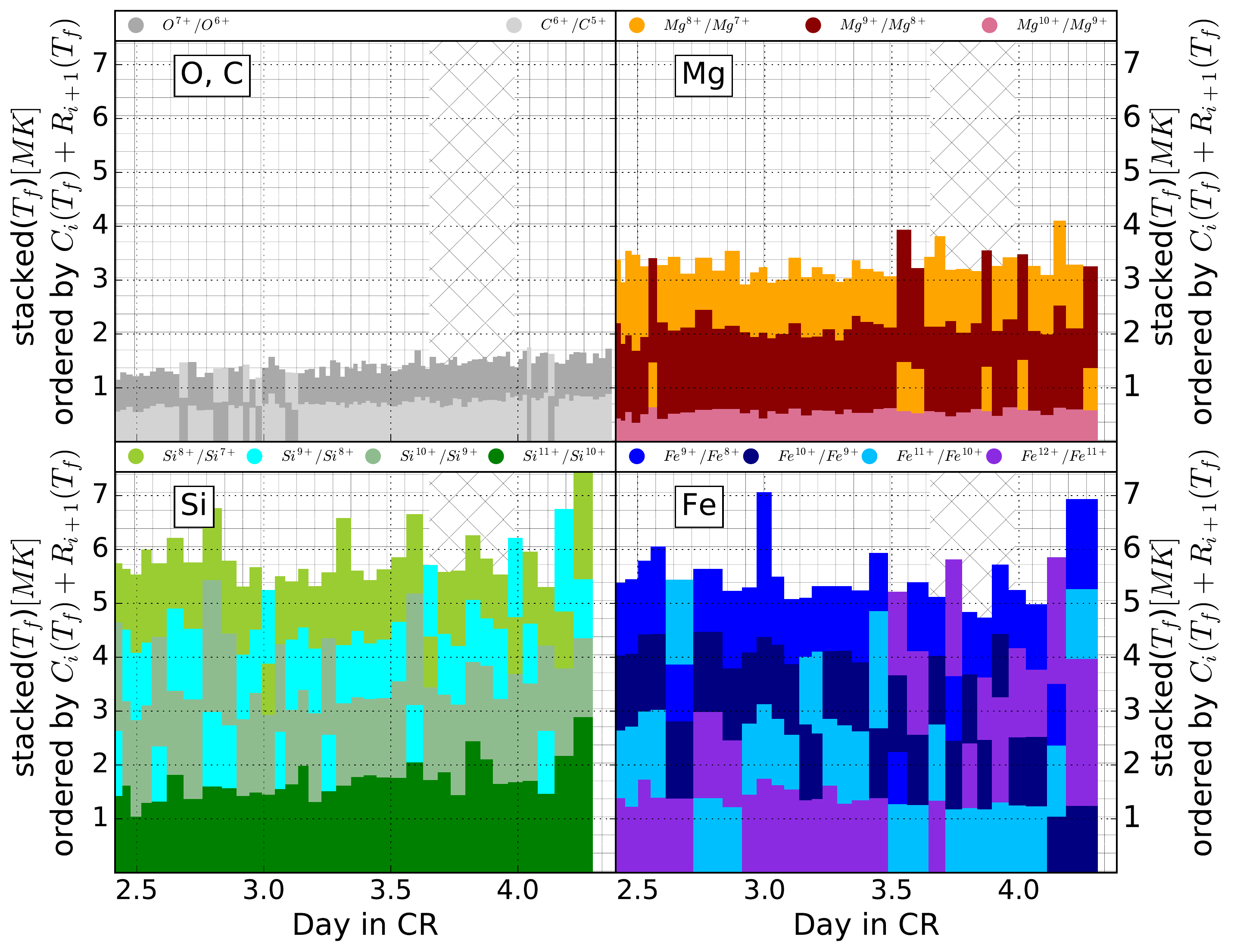}

\caption{\label{fig:TfTCR049}  Time series of minimal temperature profiles with adaptive bin size for the first recurring coronal hole wind stream in CR 2049 and in the same format as in Fig. \ref{fig:TfTCR041}. Top left: O and C ion pairs, top right: Mg ion pairs, bottom left: Si ion pairs and, bottom right: Fe ion pairs. All panels refer to a $\kappa$-function with $\kappa=10$ as the electron velocity distribution function.}
\end{figure*}

\section{A reoccurring coronal structure}\label{sec:horizontal}
We look at our structure of interest, which is mainly the first coronal hole wind stream in each CR, from two perspectives: a photospheric perspective and an in situ perspective. The PFSS model directly provides an overview on the magnetic configuration at the photospheric level. With the help of backmapping the photospheric source regions of the in situ observed solar wind can be identified. Each solar wind package originates from a different position in the photosphere. Together, the source regions of consecutive observation trace a curve in the photosphere. Thus, temporal variations in the in situ data translate to lateral variations in the photospheric source region. Therefore, the photospheric perspective provided by the PFSS model and the in situ observed solar wind parameters both contribute to a picture of the lateral structure of the source region in the photosphere. Comparing the properties of our structure of interest between different CRs gives an indication of their evolution on the timescale of CRs.

\subsection{Properties of a recurring coronal hole wind stream: photospheric perspective}
An overview over the complete heliographic maps derived with our back-mapping approach for CRs \markme{2039-2050} is given in Fig.~\ref{fig:CRmap}. Only the foot points of open field lines are shown in the heliographic map as colored dots. The color depends on the magnetic polarity as derived by the PFSS model. 
Consecutive maps differ most obviously in equatorial regions. Although their shapes and sizes vary,  several regions with open field lines appear in all considered CRs. In particular, two regions with outwards pointing magnetic polarity (green in Fig.~\ref{fig:CRmap}) are persistent through all CRs at heliographic latitude $-40^\circ<\theta<40^\circ$ and heliographic longitudes,  $40^\circ<\phi<120^\circ$ and $230^\circ<\phi<310^\circ$. With inwards pointing polarity (red in Fig.~\ref{fig:CRmap}) we find one recurring region of open field lines at heliogaphic latitude $-15^\circ<\theta<30^\circ$ and heliographic longitude $300^\circ<\phi<30^\circ$.
The shape of the border of open field line regions at the poles also changes slowly.  

In the following, we focus on the source region of a reoccurring coronal hole wind stream. The respective region is already outlined with black boxes in Fig.~\ref{fig:CRmap}. In Fig.~\ref{fig:CRmapholes}, only this region is shown. The respective open field line region is located at the western edge in the first CRs, moves further to the west and south until the center of gravity of the open field line region has reached low longitudes. Although the open field line region is present in all considered CRs, its shape and structures evolves from thin elongated structures (CRs 2039-2041) to increasingly larger regions of open field lines (CRs 2042-2048) back to small but dense regions (CRs 2049-2050). Additionally to the foot points of open field lines (in red for inwards pointing and in green for outwards-pointing polarity) the field lines that were identified as the foot points of ACE are shown in blue. The netlike, filigree structure of the open field line region in particular in CR 2041 resembles, albeit on a larger scale, the intricate connected coronal hole structures that make up the S-web.  However, unlike the S-web, which is considered to be a potential source for slow solar wind \citep[]{Antiochos2011}, in this case, clearly solar wind with coronal hole wind properties is emitted by this sparse \markme{structure.} 

\markme{Figure~\ref{fig:CRmapholes}} also shows the respective part of the corresponding MDI synoptic maps. In these synoptic maps, the magnetic field strength is coded in grayscale with negative values with high absolute value in white (for inwards pointing polarity), high positive values in black (for outwards pointing polarity) and values close to zero in gray.  In this representation, closed loops are visible as paired clusters of high magnetic field strength with opposite polarity. Since these indicate regions with closed magnetic field lines, a coronal hole as a region with open magnetic field lines  naturally cannot be in the same place. \markme{As also discussed in \citet[]{wang2010formation},} Fig.~\ref{fig:CRmapholes} shows regions with probably closed magnetic field lines in close proximity to the predicted foot points of open magnetic field lines. In most cases (except for CR 2048) the open field line region is situated between at least two such clusters of foot points of closed field lines. In particular, in CRs 2044-2048 more space is available between the active region candidates and, at the same time, the open field line region that is the source region of our coronal hole wind stream of interest is larger and the foot points are distributed more uniformly. In the other cases, the closed field line regions constrain the shape of the likely coronal hole more strongly. CR 2049 is an interesting special case. Compared to the previous CR, a new closed field line region has appeared within the previously open field line region. As a result, the open field lines in CR 2049 are very dense and are mapped very close to this active region. An additional complication is caused by the location of this region. It is located around zero degrees longitude. Thus, it sits at the interface between the oldest and the newest observations that contributed to the underlying MDI synoptic map. Therefore, the PFSS model here is  particularly vulnerable to the effect of dynamic changes. That open field lines apparently emerge very close or even from within a closed loop region is probably both a signature of the dynamic change that occurred here and the limitations of the PFSS model. The emergence of this active region which (if it is indeed the same one) is larger in CR 2050 is a possible reason for the disappearance of the coronal hole wind stream of interest in CR 2050.  \markme{We have verified that (with the exception of CR 2049) the open field line regions predicted by the PFSS model match reasonably well with dark regions visible in the synoptic maps from the  Extreme ultraviolet Imaging Telescope \markme{\citep[EIT, ][]{delaboudiniere1995eit} on SOHO.}}
In all cases, the shape of the open field line region is constrained by the proximity to closed magnetic field line loops and the evolution of the shape of the coronal hole can be considered a result of changes that are due to more dynamic phenomena such as active regions.

Figure~\ref{fig:CRmapholes} also shows that the open field line region associated with our coronal hole wind stream of interest is moving to the south-west. This is a result of the differential rotation of the Sun \citep[]{thompson1996differential}. Since the region of interest lies at lower latitudes than region used to determine the beginning and end of each CR, this region rotates faster and thus moves to the west and south on the heliographic map.

\subsection{Properties of a recurring coronal hole wind stream: in situ perspective}

\markme{A time series of the solar wind proton speed is given in Fig.~\ref{fig:CRvsw} for each CR. The highlighting indicates the solar wind type wherein green means slow solar wind, blue identifies coronal hole wind, and red shows ICMEs. There is a data gap at the beginning of CR 2039. Each CR contains two or three distinct coronal hole wind streams. One of these, indicated by the arrows above the first and in the last panel reoccurs in all CRs except the last one. The coronal hole wind stream moves to the left, i.e. earlier in the CR. This again illustrates the differential rotation of the source region. A total of eight ICMEs is included in the Jian ICME list for this time period.}

Figure \ref{fig:properties} summarizes the average properties of slow solar wind and coronal hole wind for CRs 2039-2050. Each parameter is given as the median of all 4h time resolution data points of the corresponding wind type. In addition, the \markme{15.9th and 84.1th} percentile are provided as error bars to indicate the variability of each property. In the case of the Fe charge state, we first compute the average Fe charge state  $\tilde{q}_{Fe} = \sum_{c=7}^{13} c n_{Fe^{c+}}/ \sum_{c=7}^{13} n_{Fe^{c+}}$ and then use the median to obtain a temporal averaging.
For all solar wind parameters the variability of the 4h observations is high. The variability of the proton temperature $T_p$ is higher in coronal hole wind than in slow solar wind. For the proton density $n_p$, $n_{O^{7+}}/n_{O^{6+}}$, and $n_{C^{6+}}/n_{C^{5+}}$ the variability is larger in slow solar wind. This is not surprising and well known \citep[]{bame1977evidence,mccomas2000solar,dasso2005anisotropy}. Although the temporal median of the average Fe charge state is lower in coronal hole wind than in slow solar wind, the temporal median of the average Fe charge state for coronal hole wind is within one standard deviation of that of slow solar wind. In coronal hole wind the average Fe charge state changes depending on whether Fe-hot or Fe-cool coronal hole wind is dominant. The median of the O charge state ratio remains remarkably constant during CRs 2044-2048. For CRs 2045-2048 the median of the C charge state ratio behaves similarly.  This coincides with the CRs that are least restrained in their shape by surrounding regions with closed magnetic field lines. For all other CRs the median of the O and C charge state ratios are more variable. However, the variability of the observations as indicated by the error bars is large for all CRs. Nevertheless, this could indicate that the coronal hole wind stream in CRs 2045-2048 represents undisturbed coronal hole wind whereas, in the other cases, not only the shape of the coronal hole is affected by the sourrounding active regions, but also in the same way, the composition. This could be caused by reconnection that mixes small amounts of active region plasma into the coronal hole wind stream. Another possibility is that because of the constraints on the open field lines, individual fluxtubes interact more strongly, which could lead to a more variable solar wind plasma.

Figure~\ref{fig:CRFe} shows an overview on the average Fe charge state for each CR. The average Fe charge state is variable in all considered solar wind types. As described in \citet[]{heidrich2016observations}, the Fe charge states in coronal hole wind can be as high as in slow solar wind. \markme{This motivated the distinction between Fe-cool coronal hole wind with average Fe charge states smaller than those in most slow solar wind observations and Fe-hot coronal hole wind with average Fe charge states comparable to those observed in slow solar wind.}
Periods with Fe-hot (red, $+$-hatched) and Fe-cool (black, $x$-hatched) coronal hole wind are highlighted. Several of the coronal hole wind streams shown here exhibit at least one transition between Fe-hot and Fe-cool coronal hole wind. 
The first coronal hole wind stream  in each CR persists from CR \markme{2039-2049} but is not present anymore in CR 2050. Since this stream recurs most frequently and exhibits at least one transition from Fe-hot to Fe-cool coronal hole wind in CR 2040-2049, as is shown in Fig.~\ref{fig:CRFe}, we focus on this recurring coronal hole wind stream in the following. While for CRs 2039-2041, 2043, 2044, 2047, and 2048 first a Fe-cool and then a Fe-hot coronal hole wind is observed, for CR 2042, 2045, 2046, and 2049 this is reversed. In some cases this switches back and forth.
The end of the first coronal hole wind stream in CR 2040 shows a distinct peak in the average Fe charge state. This could be a signature of a hidden ICME that is not included in the Jian ICME list. 

\section{Implications for temperature profiles}\label{sec:vertical}
Figure~\ref{fig:CSCompFe} shows the average charge state distributions for the Fe-hot and Fe-cool parts of the recurring coronal hole wind stream for each CR for Fe. Figure~\ref{fig:CSCompSi} shows the same for Si. As reference the median of the average Fe charge state of all coronal hole wind is shown as a black vertical line, its variability is indicated with gray shading, the median of the  average Fe charge state in Fe-hot coronal wind is shown with a red vertical line and for Fe-cool coronal hole wind with a \markmenew{blue} vertical line. 
This illustrates that the Fe charge state distribution is consistently shifted to higher charge states in Fe-hot coronal hole wind compared to Fe-cool coronal hole wind.  For Si this is not necessarily the case. For example for CRs 2042, 2043, 2047, and 2049 the complete Si charge state distribution is shifted to higher charge states in Fe-hot wind compared to Fe-cool wind as well. However, for the remaining CRs the Si charge state distribution in Fe-hot wind is similar to that in Fe-cool wind. Since this is coronal hole wind, the corresponding streams are all O and C cool at the same time.

In the freeze-in scenario, the charge states of two neighboring ions can change freely in the corona through recombination and ionization processes until the charge modification timescale $\tau_{\text{mod},i}(T) = \frac{1}{n_e(C_i+R_{i+1})}$ is of the same order as the expansion time scale $\tau_{\text{exp}}= \frac{H}{v_p}$. Here, $C_i$ denotes the ionization rate of the $i$th charge state, $R_{i+1}$ the recombination rate from charge state $i+1$ to $i$, $n_e$ the electron density, $H$ the electron density scale height and $v_p$ the proton bulk solar wind speed. After this point is reached, the electron density is too small and recombination does not have a significant effect anymore. Thus, the charge states of the considered ion pair is frozen-in and remains unchanged. Under the assumption of ionization equilibrium the electron temperature at the freeze-in point ($T_f$, also called freeze-in temperature) can be \markme{estimated} from the in situ observed ion densities, $n_{i}$ for the density of the $i$th charge state and $n_{i+1}$ for the $(i+1)$th charge state:
\begin{equation}\label{eq:tfeq}
n_{i}/n_{i+1}= R_{i+1}(T_f)/C_i(T_f)
.\end{equation}
Thus, in this model, pairs of charge states freeze in at different temperatures. However, this simple point of view can be misleading. If, for example, the ratio $n_{Fe^{11+}}/n_{Fe^{10+}}$ is already frozen in but $n_{Fe^{10+}}/n_{Fe^{9+}}$ has not yet reached its freeze-in point the density $n_{Fe^{10+}}$ can still change. This would also affect  the already frozen-in ratio $n_{Fe^{11+}}/n_{Fe^{10+}}$ unless the transition $Fe^{11+} \leftrightarrow Fe^{10+}$ continues to be in ionization equilibrium until its neighboring charge state pairs have frozen in as well. 

From the in situ observations, only freeze-in temperatures can be inferred directly but not the radial position of the freeze-in point. For this, a model of the solar corona is required. A variety of different models with varying complexity is available. For example, a qualitative electron temperature profile with unspecified radial dependence is derived from the in situ observed freeze-in temperature in \citet[]{geiss1995}. The declining slope of the electron temperature profile, as well as the electron density and solar wind speed in the same region, are approximated with power laws for slow solar wind in \markme{\citet[]{aellig1997}}. The coronal model of \citet[]{ko1997empirical} also derives a simple model for the electron temperature, density, and solar wind speed, but covers a larger part of the electron temperature profile, including the coronal maximum of the electron temperature. The self-consistent model of the solar corona described in \citet[]{cranmer2007self} of course also includes a detailed temperature profile. However, these models differ in the shape, maximum value, and radial position of the maximum of the electron temperature profile.
Unfortunately, it is difficult to verify which of these models describes the conditions in the solar atmosphere most accurately. This very complex issue is beyond the scope of the work presented here. Instead, we focus on the information contained directly in the in situ data and add as few assumptions as possible. A strong assumption is required to translate the in situ observed ion densities into freeze-in temperatures: the assumption of (local) ionization equilibrium. Since the ionization and recombination rates depend on it, we also have to make an assumption on the shape of the electron velocity distribution. Under thermal equilibrium, a Maxwellian distribution would be the appropriate choice. However, the solar corona is highly dynamic and thus most likely not in thermal equilibrium in the relevant regions \citep[]{contesse2004analysis,poppenhaeger2013non,singh2006spectroscopic}. Thus, a non-Maxwellian distribution is more realistic. This can be approximated for example by a $\kappa$-distribution \citep[]{dzifvcakova2015kappa,dzifvcakova2013h}. The parameter $\kappa$ controls the shape of the distribution. For $\kappa \rightarrow \infty$, the $\kappa$-distribution converges to a Maxwellian \markme{distribution.}

\markme{Figure~\ref{fig:Tfmxwkappa}} shows time series of the freeze-in temperature of four ion pairs, namely   $T_{f, Fe^{10+}/Fe^{9+}}, , T_{f, Si^{9+}/Si^{8+}}, T_{f, Mg^{9+}/Mg^{8+}}$, and $T_{f, O^{7+}/O^{6+}}$, for recombination and ionization rates based on a Maxwellian and $\kappa$-distributions with $\kappa=3$ and $\kappa=10$ as electron velocity distribution functions. The rates based on a Maxwellian electron velocity distribution function are taken from CHIANTI \citep[]{dere1997chianti,landi2013chianti} and the rates based on $\kappa$-distributions are taken from the KAPPA package for CHIANTI as described in \markme{\citet[]{dzifvcakova2015kappa}}.
While the freeze-in temperatures $T_{f, Mg^{9+}/Mg^{8+}}$ and $T_{f, Fe^{10+}/Fe^{9+}}$ are increased the further the electron velocity distribution deviates from the Maxwellian case, the opposite is the case for $T_{f, O^{7+}/O^{6+}}$. The $T_{f, Si^{9+}/Si^{8+}}$ case represents a mixture of both. For freeze-in temperatures below $1.5$ MK the most strongly superthermal electron velocity distribution function with $\kappa=3$ leads to lower freeze-in temperatures than the Maxwellian case, while for higher freeze-in temperatures the $\kappa=3$ scenario results in higher freeze-in temperatures than for a Maxwellian electron velocity distribution function.  Under model assumptions that lead to similar freeze-in radii for O and C, \citet[]{owocki1999charge} argue that both should also have comparable freeze-in temperatures. This can be achieved not only by a non-Maxwellian electron velocity distribution function  but also by taking differential streaming into account. As illustrated in Fig.~\ref{fig:Tfmxwkappa}, the different ion pairs are affected differently by different assumptions on the shape of the underlying electron velocity distribution function. Since differential streaming \citep[]{esser2001differential} is observed in the solar wind and is observed in coronal hole wind \citep[]{berger2011systematic,janitzek2016high}, a combination of both differential streaming and an extremely superthermal $\kappa$ distribution might be unrealistic. Thus, we chose $\kappa=10$ as a compromise that results in a significant, but not extreme deviation from the Maxwellian case. 

\subsection{Freeze-in order: examples based on the Cranmer 2007 model}
In Figs.~\ref{fig:profiles} and~\ref{fig:timescales}, we illustrate the freeze-in scenario with the help of the \citet[]{cranmer2007self} model. We take the temperature profile, mass density, and bulk solar wind speed for the coronal hole scenario from \citet[]{cranmer2007self}, derive the electron density with the help of the ionization fraction that is given there as well (see Fig.~\ref{fig:profiles}), and compute the expansion time scale $\tau_{exp}$, as well as the charge modification time scales $\tau_{mod}$ for selected Si and Fe ion pairs (upper subpanels in Fig~\ref{fig:timescales}). Then we modify the temperature profile to show how this affects the charge modification timescales and, in particular, the order in which charge states of the ion pairs freeze in. The recombination and ionization rates were again taken from the KAPPA package for CHIANTI and are shown in the bottom panels. In both figures, we consider four cases: the original temperature profile from \citet[]{cranmer2007self} (top left), a temperature profile with steeper slopes (top right), a down-scaled temperature profile (bottom left), and the combination of both, down-scaled temperature profile with steeper slopes (bottom right). For all three modified temperature profiles, we leave the other quantities at their original values to focus solely on the effect of the shape of temperature profile on the freeze-in order. For all four cases, the freeze-in points are indicated in Fig.~\ref{fig:timescales} as colored diamond-shaped symbols.

These scenarios only serve as illustrations of how freeze-in order can depend on the temperature profile. In the unmodified case, the Si and Fe charge state freeze-in in the expected order: The highest-order Si charge state pair $Si^{11+}/Si^{10+}$ freezes in first, followed by  $Si^{10+}/Si^{9+}$,  then $Si^{9+}/Si^{8+}$, and finally $Si^{8+}/Si^{7+}$. The same is the case for the Fe charge states: 1st $Fe^{12+}/Fe^{11+}$, 2nd $Fe^{11+}/Fe^{10+}$, 3rd $Fe^{10+}/Fe^{9+}$,  and 4th $Fe^{9+}/Fe^{8+}$. If the temperature profile has steeper slopes (top right in Figs.~\ref{fig:profiles} and~\ref{fig:timescales}) the Fe ion pairs still freeze in in the same order, but for Si the freeze-in order has changed to: 1st $Si^{11+}/Si^{10+}$, 2nd $Si^{8+}/Si^{7+}$, 3rd $Si^{9+}/Si^{8+}$,  and 4th $Si^{10+}/Si^{9+}$. The lower subpanels in Fig. \ref{fig:timescales} show the sum of the ionization and recombination rates as solid lines for each ion pair, the ionization rate as dashed lines, and the recombination rate as dotted lines. The order reversal is caused by the more localized transition from a recombination-dominated charge modification timescale at lower temperatures to an ionization-dominated charge modification timescale at higher temperatures. The only other factor in the charge modification timescale is the electron density, which is kept unchanged between the different scenarios. Thus, at their freeze-in point, some ion pairs are still dominated by recombination, while others are ionization-dominated at their respective freeze-in points. This leads to a variation of the freeze-in order. If, instead, the temperature profile is down-scaled, the freeze-in order of the Si ion pairs remains unchanged, but three of the Fe ion pair charge states,  $Fe^{11+}/Fe^{10+}$, $Fe^{10+}/Fe^{9+}$,  and $Fe^{9+}/Fe^{8+}$, freeze in very close together, with very similar freeze-in radii and freeze-in temperatures. This means that the apparent order in which they freeze-in is determined by small fluctuations and would be expected to vary frequently in the in situ derived freeze-in temperatures. We note that the assumption that each ion pair remains in ionization equilibrium at least until its neighboring ion pairs have also frozen in is, in this case, fulfilled intrinsically for $Fe^{10+}/Fe^{9+}$ because its neighbors freeze-in at the same place. The last case, in the bottom right, is the combination of both effects: the temperature profile is down-scaled and steeper compared to the original one. Here, the freeze-in order of the Si ion pairs is changed to: 1st $Si^{8+}/Si^{7+}$, 2nd $Si^{11+}/Si^{10+}$, 3rd $Si^{9+}/Si^{8+}$,  and 4th $Si^{10+}/Si^{9+}$. At the same time, while $Fe^{12+}/Fe^{11+}$ still freezes in first, the other three Fe ion pairs again freeze in very close together, i.e. their freeze-in order is just about to change.  

\subsection{Minimal temperature profiles}
Before we can compare the in situ derived freeze-in temperatures and freeze-in orders, a concise representation of this data is needed. We derive minimal temperature profiles as defined in the following. The aim of this representation is to exploit all information that is provided by the in situ observed ion density data without relying on models for the electron temperature, electron density, and solar wind speed in the relevant regions of the Sun. As long as the underlying assumption required to derive freeze-in temperatures is valid, that is that the local environment in which the ion pairs freezes in is approximately in local ionization equilibrium and that the electron velocity distribution function is well-represented by a $\kappa$-distribution, models for electron temperature, density, and solar wind speed profiles should be consistent with the observations shown here.
From in situ observations of ACE/SWICS, the densities of the most several O, C, Mg, Si, and Fe ions are available. With Eq. \ref{eq:tfeq} and the recombination and ionization rates from CHIANTI and the KAPPA package, freeze-in temperatures can be inferred for each pair of adjacent ions. We estimate the freeze-in order based on the sum of recombination and ionization rates at the respective freeze-in temperatures. In this way, the recombination and ionization rates at the freeze-in temperature additionally provide the order in which the ion pairs freeze in (see also Fig.~\ref{fig:timescales}). \markme{Figure~\ref{fig:tfmedian} shows a so-called minimal temperature profile averaged over one coronal hole wind stream (the first in CR 2045). The x-axis gives the averaged freeze-in order for 13 O, C, Mg, Si, and Fe ion pairs from low (left) to high (right) in the solar atmosphere as defined by the sum of the respective recombination and ionization rates. From models like the \citet[]{cranmer2007self} model a single maximum in this minimal temperature profile would be expected. The averaged minimal temperature profile in Fig.~\ref{fig:tfmedian} shows an additional local minimum for $T_{f, Mg^{8+}/Mg^{7+}}$. This is probably caused by the combination of two effects: 1) As a result of the variability in the freeze-in order, averaging the freeze-in order can be misleading. This is therefore avoided in the following. 2) In ACE/SWICS ions with a similar mass-to-charge ratio can influence each other. We believe that this is the reason that the densities $n_{Mg^{8+}}$ and $n_{Mg^{10+}}$ are systematically underestimated and therefore the freeze-in temperatures $T_{f, Mg^{10+}/Mg^{9+}}$ and $T_{f, Mg^{8+}/Mg^{7+}}$ are also underestimated , while $T_{f, Mg^{9+}/Mg^{8+}}$ is overestimated.}%

Figure~\ref{fig:TfTCR} combines all in situ available information into a time series plot. Each color in Fig.~\ref{fig:TfTCR} identifies an ion pair. For each temporal bin, a minimal temperature profile is \markme{shown} by stacking the freeze-in temperatures for the considered ion pairs on top of each other. They are \markme{again} ordered by the sum of their recombination and ionization rates (at the respective temperature) from lowest (and thus deepest in the solar atmosphere) to highest (which corresponds to a freeze-in radius higher in the solar atmosphere). In this way, each stack approximates a temperature profile from lowest in the atmosphere (bottom) to highest above the atmosphere (top) without giving the actual radial distance. The location of each freeze-in point requires some kind of model of the corona \markme{\citep[as for example][]{aellig1997,geiss1995,cranmer2007self}}. We now avoid using an additional model and focus on conclusions based directly on the in situ observations. To ensure comparable statistics in each bin, a variable bin size is used. For each bin and beginning with the native 12-min resolution of ACE/SWICS, the bin size is increased until at least 100 total counts per element are reached. The average bin size in all minimal temperature profiles shown here lies between 1h and 1.5h and the maximum allowed bin size of 4h is used for at most 20 bins in all the following figures.

Figure~\ref{fig:TfTCR} shows a time series of such minimal temperature profiles  for CRs 2041, 2045, and 2049. These three CRs were chosen as representatives of coronal hole streams that are mapped back to a thin sparse open field line region (CR 2041, see Fig.~\ref{fig:CRmapholes}), to an open field line region with foot points distributed more uniformly over a larger area (CR 2045), and the special case of very dense and compact open field line region very close to an active region (CR 2049). In the background, the solar wind type is indicated in the same way as in Fig. \ref{fig:CRvsw} and Fig.~\ref{fig:CRFe}. \markme{Although not shown here, we  compared} the time series of minimal temperature profiles based on a Maxwellian distribution with that based on a $\kappa$-distribution with $\kappa=10$. \markme{We  verified that the} resulting minimal temperature profiles for the Maxwellian and $\kappa$ cases are similar but not identical. In particular, both show comparable changes in the freeze-in order.

Keeping a simplified sketch of the expected shape of the electron temperature profile in mind, for example as in Fig. 5 in \citet[]{geiss1995}, helps to interpret our minimal temperature profiles. O and C ion pairs are expected to freeze in lower in the corona, in particular below the maximum of the electron temperature profile. Mg and Si ion pairs have freeze-in points around the maximum and Fe ion pairs are expected to freeze-in at higher distances and thus on the declining slope of the temperature profile.  It is notable that the example from the \citet[]{cranmer2007self} model in Fig.~\ref{fig:profiles} differs from the idealized sketch in so far as all ion pairs freeze in on the increasing slope of the temperature profile, whereas in the ACE/SWICS data (as is discussed in the following) the maximum freeze-in temperature is typically observed for $T_{f, Si^{11+}/Si^{10+}}$.
Smaller O and C freeze-in temperatures indicate a steeper increase towards the maximum of the temperature profile. The Si and Mg freeze-in temperatures are assumed to be distributed around the maximum of the temperature profile and thus estimate the maximum height of the profile. The Fe ion pairs freeze-in on the declining slope of the temperature profile and for example high Fe freeze-in temperatures indicate therefore a slowly decreasing slope of the temperature profile outside of the maximum. 

Mainly because of the smaller O and C freeze-in temperatures, the sum of all temperatures in coronal hole wind in Fig.~\ref{fig:TfTCR} is smaller than in slow solar wind. This reflects the different conditions in the respective source regions. In all three selcted CRs, the minimal temperature profile of slow solar shows high variability.

Figures~\ref{fig:TfTCR041}, \ref{fig:TfTCR045}, and \ref{fig:TfTCR049} have the same format as Fig.~\ref{fig:TfTCR}, but show a smaller subset of ion pairs and zoom in on the time period of our coronal hole wind stream of interest.  The bin size is again adaptive and therefore differs between the different subplots. Fe-hot and Fe-cool coronal hole wind streams are marked with $+$-shaped hatching (Fe-hot) and $x$-shaped hatching (Fe-cool).
The y-axis scale is, for all three figures, the same in all subplots. The comparison between the summed freeze-in temperatures for the different elements shows that the Si ion pairs have (most of the time) higher freeze-in temperatures than the Fe ion pairs. In particular in most bins, $T_{f, Si^{11+}/Si^{10+}}$ shows the highest observed freeze-in temperature. Thus, the Si ion pairs are indeed most representative of the maximum of the temperature profile. The combined Fe freeze-in temperatures are however similar. This can indicate a slow decline of the temperature profile in the corona. 
The inner part of the temperature profile, as estimated by the O and C freeze-in temperatures, shows little variability for CR 2045 but, in CRs 2041 and 2049, the O and C freeze-in temperatures and, in particular, the freeze-in order are more variable. In all three CRs in some time bins, the ratio $n_{O^{7+}}/n_{O^{6+}}$ freezes in lower in the corona than $n_{C^{6+}}/n_{C^{5+}}$. The Mg ion pairs show no difference between the Fe-hot and Fe-cool coronal hole wind streams in all three CRs. The freeze-in order of $Mg^{9+}/Mg^{8+}$ and $Mg^{8+}/Mg^{7+}$ change more frequently in CR 2041 and CR 2045 than in CR 2049. The Si ion pairs that are closest to the maximum of the temperature profile is more variable in CRs 2041 and 2049. In particular, the order in which the Si ion pairs freeze in changes frequently. Because, most of the time, they have similar freeze-in temperatures, this can also indicate that these charge states freeze-in in close radial proximity. There are a few interesting cases (for example for the Si ion pairs at DoY 78.67 and for the Fe ions at DoY 79.90) where the minimal temperature profiles exhibit more than one local maximum. Although the adaptive bin-size reduces the effects of varying statistics this could still be an effect of insufficient statistics for some of the relevant ion pairs. But it can also be interpreted as an indication for a more complex structure within the corona. The Fe-part of the temperature profile, which probably represents the declining slope at higher radial distances, is also variable, both in terms of freeze-in temperatures and in terms of the order in which the Fe ion pairs freeze in. In particular, while in Fe-cool coronal hole wind $n_{Fe^{12+}}/n_{Fe^{11+}}$ ratio (purple in Figs.~\ref{fig:TfTCR041}-\ref{fig:TfTCR049}) freezes as the first Fe ion pair in some cases, in others this is the last Fe ion pair to freeze-in (at least of those shown here). However, this correlates only weekly with Fe-hot and Fe-cool coronal hole wind. Thus, although the average Fe charge state is different in Fe-cool and in Fe-hot coronal hole wind, this is not clearly reflected in the respective minimal temperature profiles. In CR 2045, $n_{Fe^{12+}}/n_{Fe^{11+}}$ ratio freezes in first both in Fe-hot and in Fe-cool wind, in CRs 2041 and 2049 this ion pair freezes in at higher radial distances than the other Fe ion pairs mainly in Fe-hot coronal hole wind. In all three CRs, the Fe freeze-in temperatures are similar for all Fe ion pairs. If this was solely influenced by the form of the temperature profile, according to the considerations based on Fig.~\ref{fig:timescales}, this hints at a locally comparatively low electron temperature at the freeze-in point of these Fe ion pairs. 
In the case of CR 2045 the minimal temperature profile shows less variability than in CR 2041 and CR 2049. Although not shown here, this is also the case for CRs 2044, and 2046-2048. This supports the observation that our coronal hole wind stream of interest in CR 2045 as an example for an open field line region that is less strongly constrained by surrounding active regions shows undisturbed and quiet coronal hole wind.

\section{Discussion and conclusions}\label{sec:conclusions}
We have combined in situ solar wind observations from ACE with a ballistic back-mapping and a numerical PFSS to trace the evolution of a coronal hole structure and the coronal hole wind stream originating there over 11 CRs in 2006 (in CR 2050 the coronal hole wind stream is not observed in situ anymore). 
We have shown that the equatorial coronal hole shows high variability on small scales, both with respect to its shape and to its Fe charge state composition. In each CR, the in situ observations trace a different lateral path through the coronal hole.  Therefore, although solar wind from the same recurring coronal hole is observed multiple times, its properties can be expected to vary even if the source region itself would remain unchanged. 
As illustrated by the PFSS model output, the shape of the open field line region varies from CR to CR. This is probably influenced by the more dynamic neighboring coronal loop structures. These closed field line regions constrain the available space the open field line region can extend to. Interestingly, during the CRs where the open field line region is most uniformly filled with field lines, also the median O and C charge state ratios observed in situ are mostly similar. This could imply that in the CRs where the surrounding regions with closed field lines impose stronger constraints on the shape of the open field line region, their proximity also influences the compositional properties. This could be caused by reconnection effects or by inducing stronger interactions between neighboring individual flux tubes within the coronal hole wind stream. 

The average Fe charge state exhibits an interesting variability within this recurring coronal hole wind stream. Except for the case of CR 2050, where the coronal hole wind stream is not observed in situ any more, each CR shows transitions between Fe-hot and Fe-cool coronal hole wind. But their order (from Fe-hot to Fe-cool or vice versa) and length is different between CRs. This illustrates that a (recurring) coronal hole wind stream carries additional fine structure features. It is possible that the Fe-hot and Fe-cool property of a coronal hole wind stream is caused or enhanced by wave-plasma interactions. This topic requires further investigation that is beyond the scope of this study. %

We probed the coronal temperature profile based on in situ derived freeze-in temperatures. We illustrated that the freeze-in order of Si and Fe ion pairs can change depending on the respective influence of ionization versus recombination at the freeze-in point. Together with the freeze-in order determined by the sum of ionization and recombination rates the freeze-in temperatures constitute what we call a minimal temperature profile. Minimal temperature profiles can be derived without a model of the solar atmosphere. Although the radial position of the freeze-in points is not defined by the minimal temperature profiles, they nevertheless constrain the shape of the coronal electron temperature profile. We found signatures of variability of the temperature profile within coronal holes. While, in the inner part of the electron temperature profile and thus in the lower atmosphere, the shape of the temperature profile shows only small variations within the considered coronal hole wind streams, this is different higher in the corona. The probable maximum value of the electron temperature profile can change on timescales of hours, even in coronal hole wind. The declining slope of the electron temperature profile is variable as well.  This implies that the conditions in the solar corona also change on relatively small scales within a coronal hole wind stream. This variability in the electron temperature profile could be a signature of individual flux tubes. Whether it is related to turbulence remains to be investigated. The order in which the different ion pairs freeze in is here determined by the combination and ionization rates at the derived freeze-in temperature. It is sensitive to the precise conditions in the coronal hole wind stream. For models that are compatible with the assumptions necessary to derive our minimal temperature profiles, that is local ionization equilibrium and Maxwellian or $\kappa$-distributions for the underlying electron velocity distribution functions, our minimal temperature profiles can be a suitable tool for the comparison with model predictions and thus help to distinguish between different such models. 

The variability and lack thereof of the coronal hole wind properties and of the shape of the corresponding open field line region are tracers of the lateral structure of a coronal hole within the photosphere. The minimal temperature profile based on the charge state composition complements this with a probe of the radial structure of the coronal hole. For CRs 2045-2048, both the lateral and radial variability between CRs is lower than for the other CRs. This indicates that during these CRs 2045-2048, the respective coronal hole wind stream probably represents undisturbed coronal hole wind.

\begin{acknowledgements}
      Part of this work was supported by the 
      Deut\-sche For\-schungs\-ge\-mein\-schaft (DFG)\/ project
      number  Wi-2139/10-1\enspace. \markme{We are very grateful for the detailed and helpful suggestions provided by the anonymous referee.}

\end{acknowledgements}

%
   \bibliographystyle{aa} 
   \bibliography{aa} 
%
\end{document}